\documentclass[amsmath,amssymb,superscriptaddress,nobalancelastpage,prb,twocolumn]{revtex4-1}

\usepackage{graphicx}
\usepackage{varioref}
\usepackage{xr-hyper}
\usepackage{xcolor}
\usepackage{hyperref}
\hypersetup{colorlinks,linkcolor=blue,urlcolor=blue,citecolor=blue}
\usepackage{ulem}
\usepackage{braket}

\newcommand{\LSCO}{La$_{2-x}$Sr$_x$CuO$_4$}
\newcommand{\SLSCO}{SmLa$_{1-x}$Sr$_x$CuO$_4$}
\newcommand{\LESCO}{La$_{1-x/2}$Eu$_{1-x/2}$Sr$_x$CuO$_4$}
\newcommand{\Hg}{HgBa$_2$CuO$_{4+\delta}$}

\newcommand{\Tc}{$T_\mathrm{c}$}
\newcommand{\EF}{$E_\mathrm{F}$}

\newcommand{\dz}{$d_{3z^2-r^2}$}
\newcommand{\dx}{$d_{x^2-y^2}$}

\begin{document}
\author{M.~Horio}
	\email{mhorio@issp.u-tokyo.ac.jp}
 	\affiliation{Institute for Solid State Physics, The University of Tokyo, Kashiwa, Chiba 277-8581, Japan}

\author{X.~Peiao}
	\affiliation{Institute for Materials Research, Tohoku University, Katahira, Sendai 980-8577, Japan}

\author{M.~Miyamoto}
	\affiliation{Institute for Solid State Physics, The University of Tokyo, Kashiwa, Chiba 277-8581, Japan}
	
\author{T.~Wada}
	\affiliation{Institute for Solid State Physics, The University of Tokyo, Kashiwa, Chiba 277-8581, Japan}
	
\author{K.~Isomura}
\affiliation{Institute for Materials Research, Tohoku University, Katahira, Sendai 980-8577, Japan}	

\author{J.~Osiecki}
	\affiliation{Max IV Laboratory, Lund University, Box 118, 22100 Lund, Sweden}

\author{B.~Thiagarajan}
	\affiliation{Max IV Laboratory, Lund University, Box 118, 22100 Lund, Sweden}

\author{C.~M.~Polley}
	\affiliation{Max IV Laboratory, Lund University, Box 118, 22100 Lund, Sweden}
	
\author{K.~Tanaka}
	\affiliation{UVSOR Facility, Institute for Molecular Science, Okazaki 444-8585, Japan}

\author{M.~Kitamura}
	\affiliation{Institute of Materials Structure Science, High Energy Accelerator Research Organization (KEK), Tsukuba, Ibaraki 305-0801, Japan}

\author{K.~Horiba}
	\affiliation{Institute of Materials Structure Science, High Energy Accelerator Research Organization (KEK), Tsukuba, Ibaraki 305-0801, Japan}
	\affiliation{National Institutes for Quantum Science and Technology (QST), Sendai 980-8579, Japan}
	
\author{K.~Ozawa}
	\affiliation{Institute of Materials Structure Science, High Energy Accelerator Research Organization (KEK), Tsukuba, Ibaraki 305-0801, Japan}

\author{T.~Taniguchi}
	\affiliation{Institute for Materials Research, Tohoku University, Katahira, Sendai 980-8577, Japan}

\author{M.~Fujita}
	\affiliation{Institute for Materials Research, Tohoku University, Katahira, Sendai 980-8577, Japan}

\author{I.~Matsuda}
	\affiliation{Institute for Solid State Physics, The University of Tokyo, Kashiwa, Chiba 277-8581, Japan}

\title{Influence of oxygen-coordination number on the electronic structure of	 \\ single-layer La-based cuprates}

\begin{abstract}
We present an angle-resolved photoemission spectroscopy study of the single-layer T*-type structured cuprate \SLSCO\ with unique five-fold pyramidal oxygen coordination. Upon varying oxygen content, T*-\SLSCO\ evolved from a Mott-insulating to a metallic state where the Luttinger sum rule breaks down under the assumption of a large hole-like Fermi surface. This is in contrast with the known 
doping evolution of the structural isomer \LSCO\ with six-fold octahedral coordination. In addition, quantitatively characterized Fermi surface suggests that the empirical \Tc\ rule for octahedral oxygen-coordination systems does not apply to T*-\SLSCO. The present results highlight unique properties of the T*-type cuprates possibly rooted in its oxygen coordination, and necessitate thorough investigation with careful evaluation of disorder effects. 

\end{abstract}

\maketitle

\section{Introduction}
Strong electron correlation lies at the heart of high-temperature superconductivity in cuprates~\cite{ScalapinoRMP2012}. While its major contribution to superconductivity is unquestionable, the quantitative relevance is still under debate. Recent first-principles calculations~\cite{JangSciRep2016,HirayamaPRB2018} have estimated particularly large effective onsite Coulomb interaction $U$ for one of the lowest-\Tc\ materials \LSCO (LSCO), while more modest values for the higher-\Tc\ Hg-based cuprates. It was suggested using the Hubbard model that larger $U$ for LSCO facilitates charge inhomogeneity ~\cite{MisawaPRB2014,IdoPRB2018}, which competes with superconductivity and could in turn lower \Tc~\cite{ImadaJPSJ2021}. To further gain insight into the influence of electron correlation, comparative experimental studies are desired with varying the onsite effective interaction but with minimal changes to other details such as block-layer properties.

  \begin{figure}[ht!]
  	\begin{center}
  		\includegraphics[width=0.48\textwidth]{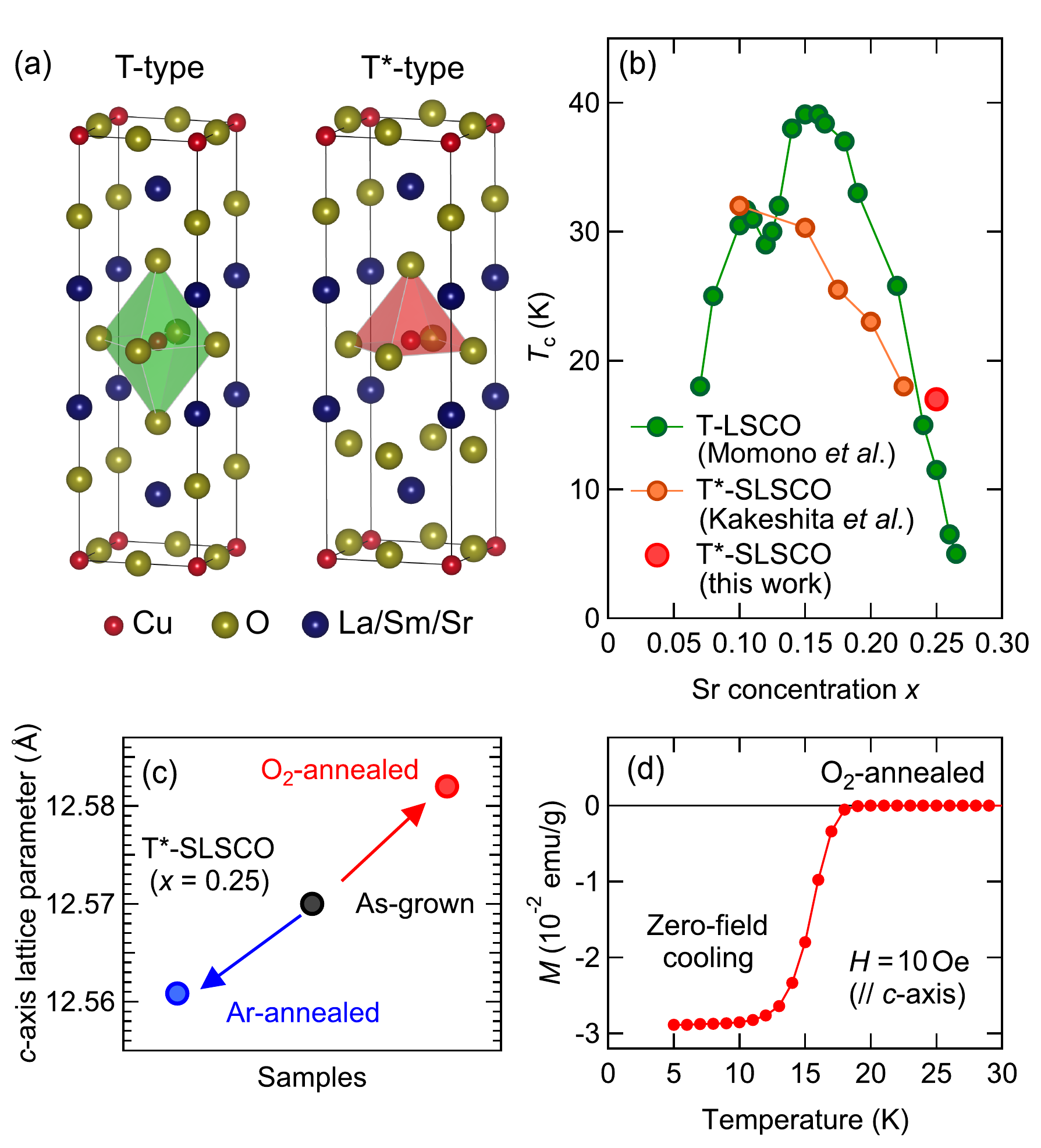}
  	\end{center}
  	\caption{\textbf{Properties of the T*-SLSCO samples.} (a) Crystal structure of the T-type (left) and T*-type (right) cuprates. (b) \Tc\ versus Sr concentration for the present O$_2$-annealed T*-SLSCO ($x = 0.25$) sample plotted along with previous results on T-LSCO~\cite{MomonoPhysC1994} and T*-SLSCO~\cite{KakeshitaJPC2009}. (c) $c$-axis lattice constant of the as-grown, Ar-annealed, and O$_2$-annealed T*-SLSCO ($x = 0.25$) samples. (d) Magnetization of the O$_2$-annealed sample as a function of temperature. }
  	\label{Phase}
  \end{figure}

\begin{figure*}[ht!]
	\begin{center}
		\includegraphics[width=0.99\textwidth]{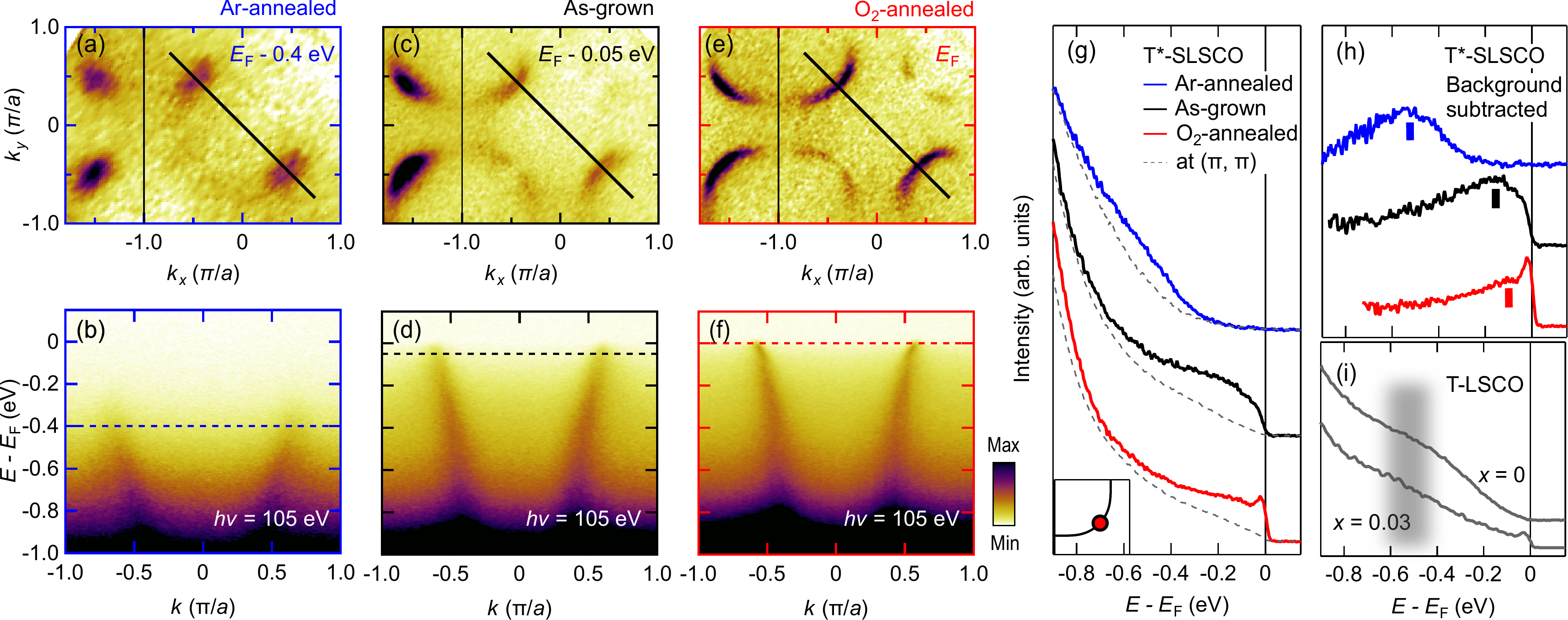}
	\end{center}
	\caption{\textbf{Evolution of the electronic structure of T*-SLSCO by annealing treatment.} (a) Constant-energy surface of the Ar-annealed sample. Intensity has been integrated over $\pm 20$~meV with respect to the indicated energy level. (b) Nodal energy-momentum map of the Ar-annealed sample. The dashed line marks the energy level where the constant-energy surface is mapped. (c)--(f) The same as (a) and (b) for the as-grown and O$_2$-annealed samples. (g) EDCs of the T*-SLSCO samples at the nodal (underlying) $k_\mathrm{F}$. EDCs around ($\pi, \pi$) are overlaid as a background. (h) EDCs after background subtraction. The LHB peak position is marked by vertical bars. The EDC at $E < E_\mathrm{F} - 0.7$~eV for the O$_2$-annealed sample is omitted due to possible contributions from other bands. (i) Nodal EDC for T-LSCO at $x = 0$ and 0.03 extracted from Ref.~\onlinecite{YoshidaJCMP07}. The black shade marks LHB.}
	\label{map}
\end{figure*}

Structural isomers of single-layer 214-type cuprates provide a unique platform for this purpose. The local oxygen coordination around Cu can be varied from six-fold octahedral [T-type, Fig.~\ref{Phase}(a)] to four-fold square-planar one (T'-type)~\cite{ArmitageRMP2010}. The reduction of the oxygen coordination number is known to cause a decrease of the charger-transfer gap and hence an increase of Cu~$3d$ -- O~$2p$ hybridization~\cite{TokuraPRB1990}, which leads to stronger screening of local interactions~\cite{JangSciRep2016,HirayamaPRB2018}. However, the T'-type cuprates do not easily accept Sr substitutions, and it is still unsettled whether they can be doped with holes~\cite{TakamatsuAPEX2012,LinJPSJ2019}. On the other hand, T*-type cuprates with five-fold pyramidal oxygen coordination [Fig.~\ref{Phase}(a)] can be hole doped and thus are directly comparable to the T-type counterpart~\cite{AkimitsuJJAP1988}. An optical conductivity study on the parent compounds revealed smaller charge-transfer gap for the T*-type cuprate than the T-type one~\cite{TokuraPRB1990}, suggesting smaller onsite effective interaction for the T*-type cuprate. It is known that as-grown samples of the T*-type cuprates have oxygen defects likely at the apical site and high-pressure post-growth O$_2$ annealing is necessary to induce superconductivity~\cite{IzumiPhysC1989,AsanoJPSJ2019}. Recent oxygen K-edge x-ray absorption (XAS) studies~\cite{AsanoJPSJ2020} found the drastic increase of doped holes by O$_2$ annealing, consistent with the compensation of the oxygen defects. The \Tc\ of O$_2$-annealed T*-\SLSCO (SLSCO) shows a monotonic increase with decreasing Sr doping~\cite{KakeshitaJPC2009}, which is in contrast to the dome-like behavior of T-LSCO~\cite{MomonoPhysC1994} [Fig.~\ref{Phase}(b)]. The distinct superconducting characteristics between the T*- and T-type cuprates suggest substantial differences in the electronic structure associated with variation of oxygen-coordination number. However, the electronic structure of T*-type cuprates including its evolution by annealing has remained largely unexplored~\cite{InoPhysB2004} even after 35 years since the discovery.

Here, we present an angle-resolved photoemission spectroscopy (ARPES) study of T*-SLSCO ($x=0.25$) annealed under various conditions. By compensating oxygen vacancies, T*-SLSCO evolves from a Mott-insulating to a metallic state in a fashion distinct from the doping evolution of T-LSCO. Furthermore, quantitative analysis of the metallic Fermi surface suggests deviation from empirical scaling law between $T_c$ and Fermi surface curvature or hole doping level. The present work suggests unique properties of the T*-type cuprates that are either intrinsic or dominated by disorder.

\section{Methods}   
Single crystals of T*-SLSCO ($x=0.25$) were grown by the traveling solvent floating-zone method. Three kinds of samples, as-grown, Ar-annealed, and O$_2$-annealed, were prepared. Ar annealing was performed at 800~$^\circ$C and 0.1~MPa for 12 hours and O$_2$ annealing at 500~$^\circ$C and 45~MPa for 72 hours. The shortening and elongation of the $c$-axis lattice constant by Ar and O$_2$ annealing [Fig.~\ref{Phase}(c)] suggest incorporation and removal of apical oxygen atoms, respectively. The change in the oxygen content was evaluated to be $\delta = -0.032$ and 0.020 after Ar and O$_2$ annealing, respectively. The O$_2$-annealed sample showed superconductivity below $T_c = 17$~K [Fig.~\ref{Phase}(d)]. This value is consistent with the known trend in the doping range of $x = 0.10-0.225$~\cite{KakeshitaJPC2009} [Fig.~\ref{Phase}(b)]. ARPES measurements were performed at the BLOCH beamline of MAX~IV at $T = 20$~K except for the Ar-annealed sample that was measured at 210~K to avoid charging. Photon energies were set at $h\nu = 105$~eV and the total energy resolution at 25~meV.

\section{Results}   
The electronic-structure evolution by annealing treatment is displayed in Fig.~\ref{map}. For the Ar-annealed T*-SLSCO ($x=0.25$) sample, the bands are entirely gapped, and a broad dispersion peaked around ($\pi/2$, $\pi/2$), resembling the lower Hubbard band (LHB) of T-La$_2$CuO$_4$~\cite{YoshidaPRL2003} and Ca$_2$CuCl$_2$O$_2$~\cite{ShenPRL2004}, is found around $\sim 0.5$~eV below $E_\mathrm{F}$ [Figs.~\ref{map}(a) and (b)]. This suggests close proximity to the Mott insulating state. On the other hand, for the as-grown sample with less oxygen vacancies, the low-energy state extends toward \EF\ [Fig.~\ref{map}(d)] and toward the antinodal direction [Fig.~\ref{map}(c)]. However, the spectral weight becomes diminishingly small with approaching \EF\ as one can see from the energy distribution curve (EDC) in Fig.~\ref{map}(g). By O$_2$ annealing and thereby further compensating apical-oxygen vacancies to induce superconductivity, a sharp nodal quasiparticle peak is created [Figs.~\ref{map}(f) and (g)]. Subtracting the EDC at ($\pi$, $\pi$) as a background, it becomes clear that the LHB is shifted toward \EF\ and the quasiparticle peak is created just above the LHB, as going from the Ar-annealed to the O$_2$-annealed samples [Fig.~\ref{map}(h)].  This evolution from the Mott insulating to superconducting states is similar to the doping evolution of other cuprates such as Ca$_{2-x}$Na$_x$CuCl$_2$O$_2$ (Na-CCOC)~\cite{KohsakaJPSJ2003,ShenPRL2004}, Bi2201~\cite{HashimotoPRB2008,PengNatCommun2013} and Bi$_2$Sr$_{2-x}$La$_x$(Ca, Y)Cu$_2$O$_{8+\delta}$ (Bi2212)~\cite{TanakaPRB2010}, where the chemical potential is continuously shifted toward LHB until a nodal quasiparticle peak is created. We thus conclude that the primary effect of oxygen-defect compensation for T*-SLSCO is to dope holes.

\begin{figure}[t!]
	\begin{center}
		\includegraphics[width=0.48\textwidth]{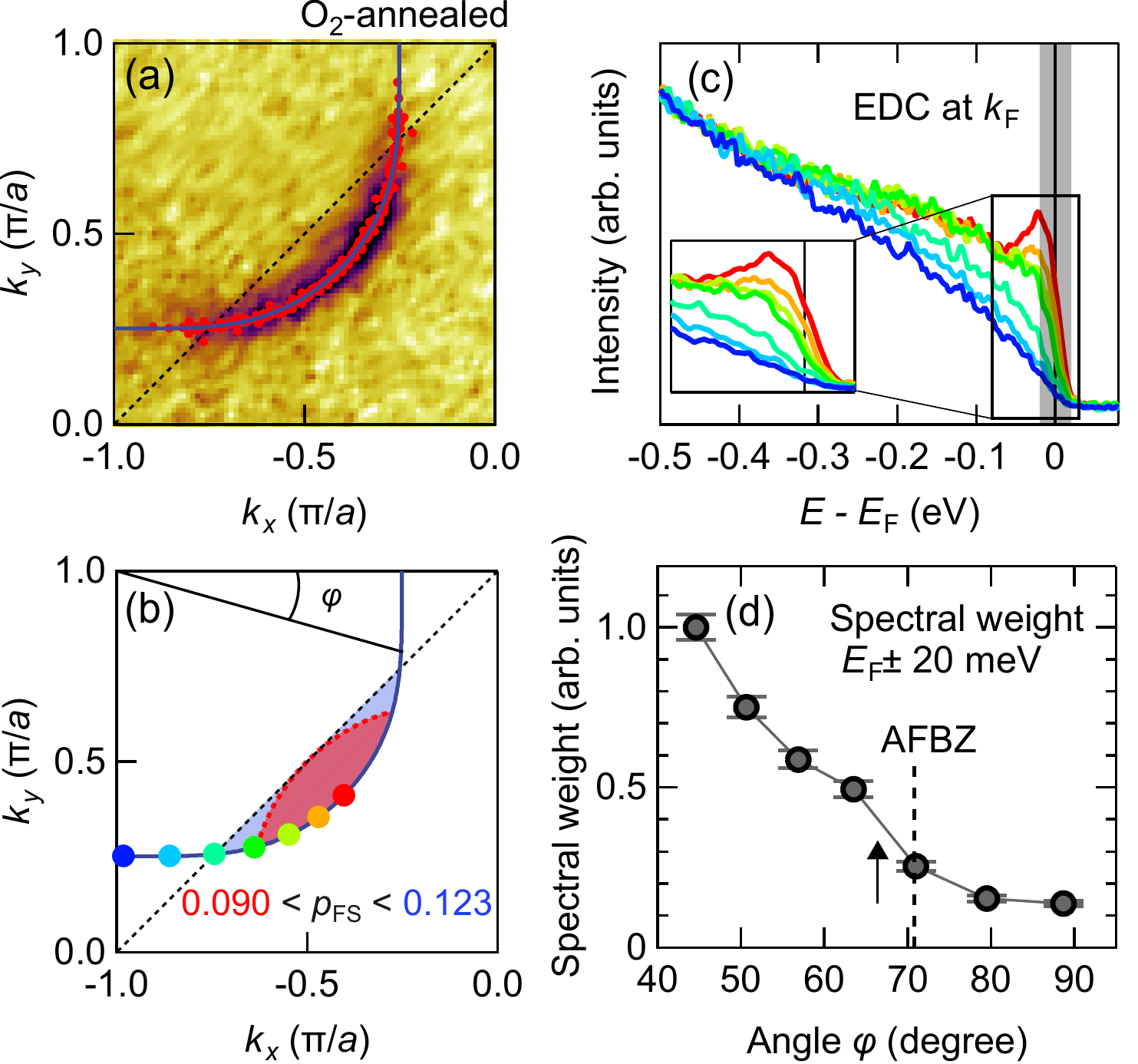}
	\end{center}
	\caption{\textbf{Fermi surface geometry.} (a),(b) Fermi surface of the O$_2$-annealed T*-SLSCO ($x=0.25$) sample fitted to the tight-binding model. Red dots in (a) represent momentum-distribution-curve (MDC) peak positions gathered into the displayed region according to the $C_4$  and mirror symmetries. Hole concentration $p_\mathrm{FS}$ is estimated from the small Fermi surface shaded in (b). See text for detailed descriptions about the small Fermi surface. (c) EDCs from the nodal to the antinodal momentum points indicated in (b). (d) Spectral weight integrated over $E_\mathrm{F} \pm 20$~meV [black shaded region in (c)] and plotted against Fermi surface angle $\varphi$ defined in (b). The error bar has been evaluated based on the uncertainty of \EF\ ($\pm 1$~meV). The arrow marks a steeper change.}
	\label{pFS}
\end{figure}

In order to quantitatively characterize the Fermi surface of O$_2$-annealed T*-SLSCO ($x = 0.25$), we fitted peak positions of the momentum distribution curves (MDCs) at \EF\ to the tight-binding model [Fig.~\ref{pFS}(a)], $\epsilon-\mu = \epsilon_0 - 2t[\cos(k_xa)+\cos(k_ya)] - 4t'\cos(k_xa)\cos(k_ya) - 2t''[\cos(2k_xa)+\cos(2k_ya)]$, where $t$, $t'$, and $t''$ represent nearest, second-nearest, and third-nearest neighbor hopping parameters, respectively, and $\epsilon_0$ is the center of the band relative to the chemical potential $\mu$. Fixing $t''/t'$ at $-0.5$ as frequently assumed~\cite{YoshidaPRB2006,HashimotoPRB2008}, we obtained $t'/t = -0.30$ and $\epsilon_0 /t = 0.54$. The validity of these tight-binding parameters have been confirmed by examining constant-energy surfaces at deeper binding energies (see Appendix~A). The area of the fitted Fermi surface in Fig.~\ref{pFS}(a) yields the doping level of 0.00. This unrealistically small value for the superconducting sample suggests breakdown of the Luttinger's theorem under the assumption of a large hole Fermi surface around ($\pi$, $\pi$). Similar violation has also been reported for underdoped Bi2201~\cite{HashimotoPRB2008} and Na-CCOC~\cite{ShenScience2005}. Yang, Rice, and Zhang proposed, based on the doped resonant
valence-bond spin-liquid concept, that the deviation implies a small Fermi surface formed within the original Fermi surface and the antiferromagnetic Brillouin-zone boundary (AFBZ) being a more appropriate description in the pseudogap state~\cite{YangPRB2006}. Subsequent ARPES studies on Bi2212 and Na-CCOC supported this picture~\cite{YangPRL2011,MengPRB2011}. Recent experimental works on the Hall coefficient of hole-doped cuprates also suggest the formation of a small Fermi surface in the presence of the pseudogap~\cite{BadouxNature2016,CollignonPRB2017,PutzkeNatPhys2021}.

 Examining EDCs at $k_\mathrm{F}$ positions [Fig.~\ref{pFS}(c)], a sharp suppression of  the Fermi edge is found as approaching AFBZ from the nodal direction. Accordingly, spectral weight integrated over $E_\mathrm{F} \pm 20$~meV exhibits a steeper reduction before reaching AFBZ [marked by the arrow in Fig.~\ref{pFS}(d)]. These observations are consistent with the small Fermi surface scenario where the Fermi surface is confined on one side of the AFBZ~\cite{YangPRB2006,YangPRL2011,MengPRB2011}. While the exact shape is difficult to elucidate, the Fermi surface area should fall between the two extreme cases~\cite{YangPRB2006,YangPRL2011,MengPRB2011}; the region enclosed by the original Fermi surface and (i) AFBZ [blue shaded in Fig.~\ref{pFS}(b)] or (ii) the mirrored Fermi surface whose nodal part is tangent to the AFBZ [red shaded in Fig.~\ref{pFS}(c)].  In order to evaluate the actual hole concentration, we estimated the doping level of each case [0.123 for case (i) and 0.090 for case (ii)] and averaged the two values. Consequently, we obtained the doping level $p_\mathrm{FS} = 0.107 \pm 0.016$. From a weight change, the difference in the oxygen content between the Ar- and O$_2$-annealed samples is $0.032+0.020=0.052$. This straightforwardly translates into the doping-level difference of 0.104 between the two samples. Considering the proximity of the Ar-annealed sample to the Mott-insulating state (Fig.~\ref{map}), the estimated  $p_\mathrm{FS}$ is consistent with the changes in the oxygen content, though being smaller than the XAS estimate ($p=0.17$) on polycrystalline O$_2$-annealed T*-\LESCO\ ($x=0.25$)~\cite{AsanoJPSJ2020}.

\section{Discussion}   
Having clarified the doping evolution of T*-SLSCO from the Mott insulating to a metallic state with a small Fermi surface, let us compare with the case of T-LSCO. For T-LSCO, upon hole doping, a quasiparticle peak emerges abruptly by spectral-weight transfer from LHB while the position of LHB remains virtually unaffected [Fig.~\ref{map}(i)]~\cite{InoPRB2000,YoshidaPRL2003,YoshidaJCMP07}. Moreover, the resulting metallic state with a large (underlying) Fermi surface obeys the Luttinger's theorem even in the underdoped region~\cite{YoshidaPRB2006}. It was proposed that the doping evolution for T-LSCO can be explained by formation of a mixed phase of antiferromagnetic and superconducting states~\cite{MayrPRB2006} with possible dynamical fluctuations~\cite{YoshidaPRB2006}. Considering that particularly large $U$~\cite{JangSciRep2016,HirayamaPRB2018} for T-LSCO could enhance a tendency toward such a phase separation~\cite{MisawaPRB2014,IdoPRB2018,CaponePRB2006}, the distinct evolution is consistent with a smaller $U$ for the T*-type cuprates expected from the smaller charge-transfer gap ~\cite{TokuraPRB1990}. 

\begin{figure}[t!]
	\begin{center}
		\includegraphics[width=0.48\textwidth]{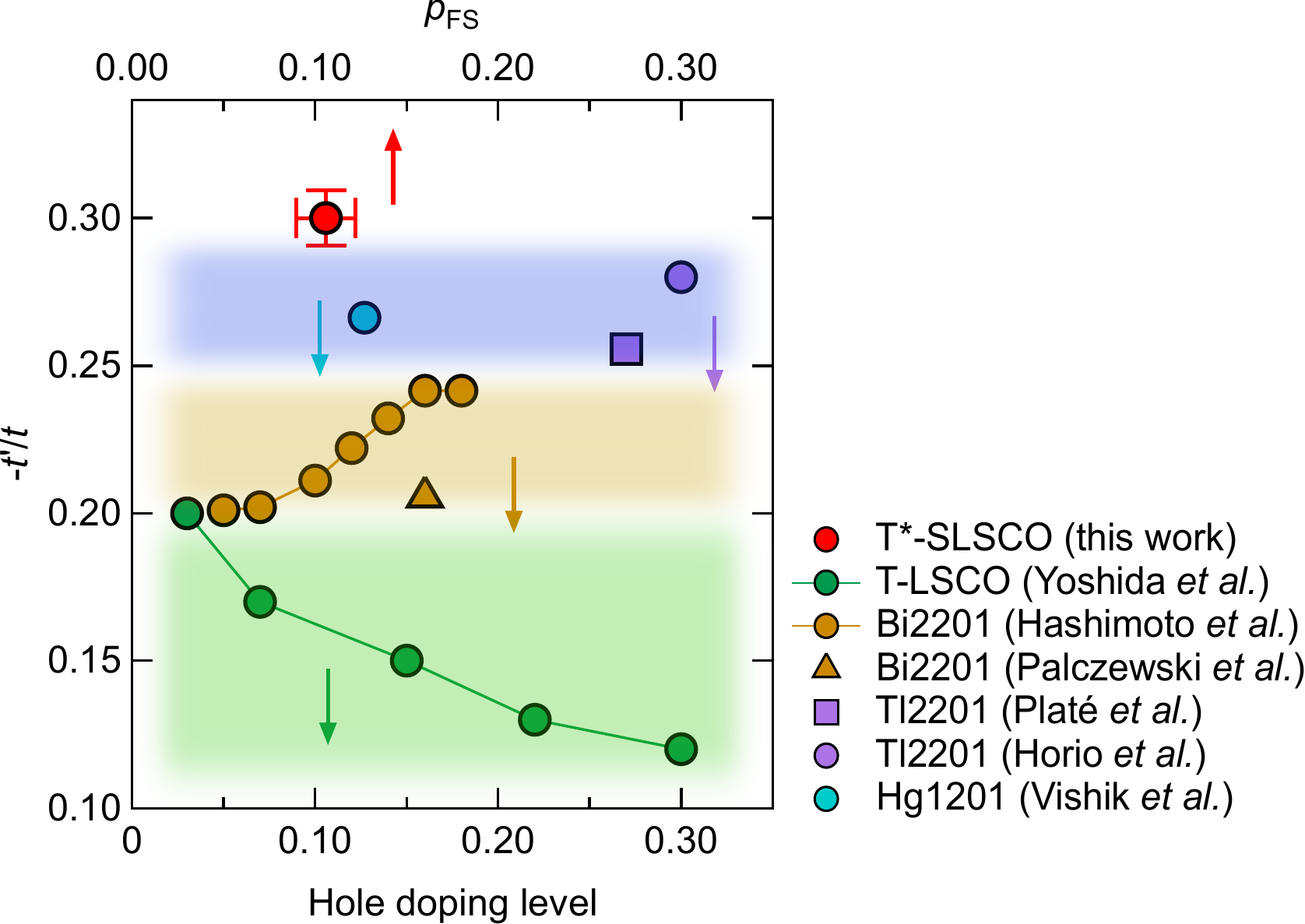}
	\end{center}
	\caption{\textbf{Hopping parameter of single-layer cuprates with octahedral or pyramidal oxygen coordination.} The value of $-t'/t$ versus hole concentration for single-layer hole-doped cuprates as indicated~\cite{YoshidaPRB2006,HashimotoPRB2008,PalczewskiPRB2008,PlatePRL2005,HorioPRL2018,VishikPRB2014}. The assumption of $-t''/t' = 0.5$ has been applied to all the data.}
	\label{hopping}
\end{figure}

In a broader perspective, it is of primary importance to make a comparison between the single-layer cuprates with pyramidal and octahedral oxygen coordination in general. To that end, the values of $-t'/t$ that defines the curvature of the underlying large Fermi surface are extracted from previous ARPES works on cuprates with octahedral oxygen coordination~\cite{YoshidaPRB2006,HashimotoPRB2008,PalczewskiPRB2008,PlatePRL2005,HorioPRL2018,VishikPRB2014} and plotted against hole doping in Fig.~\ref{hopping}. Sr concentration is employed as a doping level for T-LSCO~\cite{YoshidaPRB2006}, while for Bi2201~\cite{HashimotoPRB2008,PalczewskiPRB2008}, Tl$_2$Ba$_2$CuO$_{6+\delta}$ (Tl2201)~\cite{PlatePRL2005,HorioPRL2018}, and \Hg\ (Hg1201)~\cite{VishikPRB2014}, hole concentrations derived from transport measurements~\cite{HashimotoPRB2008,BanguraPRB2010,CuloSciPost2021,YamamotoPRB2000} are used. The present result on T*-SLSCO is plotted versus $p_\mathrm{FS}$ determined from the Fermi surface area. While $-t'/t$ somewhat depends on the doping level, the values for cuprates with octahedral oxygen coordination can be classified into three regimes based on material differences; 0.10-0.20 for LSCO, 0.20-0.25 for Bi2201, and $>0.25$ for Tl2201 and Hg1201. It has been shown that smaller values of $-t'/t$ in a single-band description result from hybridization between the \dx\ and \dz\ orbitals that pushes up the energy level of antinodal quasiparticle dispersion~\cite{MattNatCommun2018,HirayamaPRB2018}. This orbital hybridization is harmful for spin fluctuation-mediated superconductivity and thus predicted to reduce \Tc~\cite{SakakibaraPRL2010,SakakibaraPRB2012}. As such, smaller $-t'/t$ has been associated with lower \Tc\ and vice versa~\cite{PavariniPRL2001}. As for O$_2$-annealed T*-SLSCO, the $-t'/t$ value is even larger than those for Tl2201 and Hg1201 whose maximal \Tc's exceed 90~K. This is apparently incompatible with even lower \Tc\ of the T*-type cuprates than that of the T-type cuprates with exceptionally small $-t'/t$. Since the separation between the \dx\ and \dz\ energy levels is expected, from an ionic-model analysis~\cite{OhtaPRB1991}, to be comparable for the T*- and T-type cuprates, orbital hybridization is unlikely to be the main origin of the deviation. According to cellular dynamical mean-field theory~\cite{CivelliPRL2005}, $-t'/t$ is enhanced by electron correlation in the underdoped regime. However, moderate electron correlation for the T*-type cuprates would not account for the enhancement of $-t'/t$ over the values of the octahedral coordination systems, especially of T-LSCO. It is of note that Na-CCOC --  isostructural to T-LSCO but La/Sr and apical oxygen are respectively replaced by Ca/Na and Cl -- similarly possesses larger $-t'/t$ than Bi2201~\cite{HashimotoPRB2008} despite low maximal $T_\mathrm{c} \sim 27$~K. 
Although the exact mechanism remains to be addressed, it is possible that a common origin for the deviation of $-t'/t$ underlies the single-layer non-octahedral oxygen coodinated systems.

If one applies the empirical \Tc-versus-$p$ scaling commonly used for hole-doped cuprates, $1-T_c/T_{c\mathrm{,max}} = 82.6 \times (p-0.16)^2$~\cite{PreslandPhysC1991,PutzkeNatPhys2021}, the \Tc\ of the O$_2 $-annealed T*-SLSCO ($x=0.25$) 
with respect to the optimal value~\cite{KakeshitaJPC2009} yields $p = 0.235$. This would fall deep in the overdoped regime that is comparable or even beyond the typical pseudogap endpoint ($p^* \sim$~0.19--0.23)~\cite{CooperScience2009,CollignonPRB2017,PutzkeNatPhys2021}. For O$_2 $-annealed T*-SLSCO ($x=0.25$), however, the actual hole concentration amounts only to $p_\mathrm{FS}=0.107$ and a clear pseudogap is observed [Fig.~\ref{pFS}(c)]. This implies deviation from the common phase diagram; the superconducting phase is apparently condensed at low doping, and the pseudogap prevails over the larger part of the superconducting dome. In fact, the shift of the superconducting region towards low doping was predicted by variational Monte Carlo studies on the single-band Hubbard and $d$-$p$ models upon decreasing $U$~\cite{YokoyamaJPSJ2013,WatanabeJPCM2023}. On the other hand, the influence of disorder should also be considered for discussing the phase diagram. Due to small apical-oxygen height of the T*-type cuprates (2.2--2.3~\AA)~\cite{SawaNature1989,KakeshitaJPC2009} compared to those of other single-layer cuprates ($\sim 2.4$~\AA\ for T-LSCO~\cite{RadaelliPRB1994_2} and $\sim 2.7$~\AA\ for Tl2201~\cite{TorardiPRB1988}), disorder introduced by Sr substitutions are placed more closely to CuO$_2$ planes and hence would more strongly affect superconducting properties~\cite{EisakiPRB2004}. In a previous work, this disorder effect was proposed as the origin of the monotonic \Tc\ suppression upon Sr doping~\cite{KakeshitaJPC2009}. Moreover, in the present O$_2$-annealed T*-SLSCO ($x=0.25$), $x-p_\mathrm{FS}=0.143$ implies that 0.072 oxygen atoms are deficient, compensating hole doping by Sr substitutions. If the concentration of apical oxygen defects depends on Sr doping, that could also affect the doping dependence of \Tc. To establish the intrinsic electronic phase diagram of T*-type cuprates that hosts a potential uniqueness, systematic electronic structure investigation of superconducting samples is necessary in a wide doping range with a special care for the influence of disorder.

\section{Conclusions}   
In summary, we have investigated the electronic structure of T*-type cuprate SLSCO ($x=0.25$) with pyramidal oxygen coordination by means of ARPES. By compensating apical-oxygen vacancies and thus doping holes, T*-SLSCO evolved from the Mott-insulating to metallic state with a small Fermi surface. This doping evolution is similar to other hole-doped cuprates except for T-LSCO, highlighting the peculiarity of T-LSCO arising likely from its particularly large effective onsite interaction. The empirical relationships such as maximal \Tc\ versus $-t'/t$ and \Tc\ versus $p$ apparently do not hold for the T*-type cuprates. Systematic doping-dependent studies on superconducting samples with careful assessment of disorder effects will pave the way for unveiling the full picture of its potentially unique phase diagram.

\section{Appendix A: Determination of hopping parameters}
\begin{figure*}[ht!]
	\begin{center}
		\includegraphics[width=0.9\textwidth]{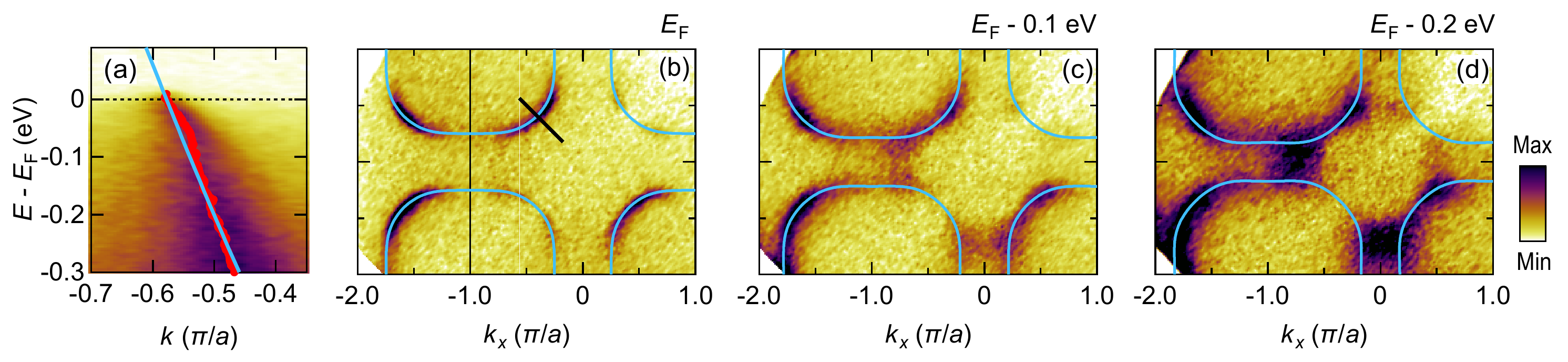}
	\end{center}
	\caption{\textbf{Tight-binding analysis at deeper binding energies} (a) Energy-momentum map along the nodal direction [indicated by black line in (b)] overlaid with MDC peak positions (red dots). The light-blue curve represents the tight-binding-modeled band dispersion with $t=0.31$~eV. (b)-(d) Constant-energy maps at indicated energy levels. The photoemission intensity has been integrated within $\pm 20$~meV for (b) and $\pm 10$~meV for (c) and (d). Constant energy surfaces modeled by the tight-binding model are plotted as light-blue curves.} 
	\label{t}
\end{figure*}

In order to test the validity of tight-binding parameters determined from the Fermi surface, we examined electronic states at deeper binding energies. Figure~\ref{t}(a) shows nodal band dispersion of the O$_2$-annealed T*-SLSCO ($x = 0.25$). Peak positions of MDCs have been extracted and plotted as red dots. From the nodal dispersion at $-0.3$~eV $<E-E_\mathrm{F}<$ $-0.15$~eV that is free from kink features,  the nearest-neighbor hopping term $t$ was determined to be 0.31~eV. Using this $t$ value, constant energies surfaces have been reproduced by the tight-binding model not only at \EF\ but also at deeper binding energies and compared with the experimental maps in Figs.~\ref{t}(b)-(d). Overall satisfactory agreement ensures the reliability of the tight-binding parameter values.

\vspace{5mm}
\begin{acknowledgments}
Fruitful discussion with T.~Yoshida, T.~Kondo, K.~P.~Kramer, and J.~Chang is greatfully acknowledged. This work was supported by JSPS KAKENHI Grant Numbers~JP21K13872 and JP21H04987. X.P. was supported by Collaborative Research Center on Energy Materials, Institute for Materials Research (IMR), Tohoku University. We acknowledge MAX IV Laboratory for time on Beamline BLOCH under Proposal 20220314. Research conducted at MAX IV, a Swedish national user facility, is supported by the Swedish Research council under contract 2018-07152, the Swedish Governmental Agency for Innovation Systems under contract 2018-04969, and Formas under contract 2019-02496. Preliminary measurements 
were also conducted at BL28A of Photon Factory (Proposal No.~2021G092) and BL5U and BL7U of UVSOR Synchrotron Facility, Institute for Molecular Science (IMS programs 22IMS6660 and 22IMS6839).
\end{acknowledgments}


\begin{thebibliography}{56}%
	\makeatletter
	\providecommand \@ifxundefined [1]{%
		\@ifx{#1\undefined}
	}%
	\providecommand \@ifnum [1]{%
		\ifnum #1\expandafter \@firstoftwo
		\else \expandafter \@secondoftwo
		\fi
	}%
	\providecommand \@ifx [1]{%
		\ifx #1\expandafter \@firstoftwo
		\else \expandafter \@secondoftwo
		\fi
	}%
	\providecommand \natexlab [1]{#1}%
	\providecommand \enquote  [1]{``#1''}%
	\providecommand \bibnamefont  [1]{#1}%
	\providecommand \bibfnamefont [1]{#1}%
	\providecommand \citenamefont [1]{#1}%
	\providecommand \href@noop [0]{\@secondoftwo}%
	\providecommand \href [0]{\begingroup \@sanitize@url \@href}%
	\providecommand \@href[1]{\@@startlink{#1}\@@href}%
	\providecommand \@@href[1]{\endgroup#1\@@endlink}%
	\providecommand \@sanitize@url [0]{\catcode `\\12\catcode `\$12\catcode
		`\&12\catcode `\#12\catcode `\^12\catcode `\_12\catcode `\%12\relax}%
	\providecommand \@@startlink[1]{}%
	\providecommand \@@endlink[0]{}%
	\providecommand \url  [0]{\begingroup\@sanitize@url \@url }%
	\providecommand \@url [1]{\endgroup\@href {#1}{\urlprefix }}%
	\providecommand \urlprefix  [0]{URL }%
	\providecommand \Eprint [0]{\href }%
	\providecommand \doibase [0]{http://dx.doi.org/}%
	\providecommand \selectlanguage [0]{\@gobble}%
	\providecommand \bibinfo  [0]{\@secondoftwo}%
	\providecommand \bibfield  [0]{\@secondoftwo}%
	\providecommand \translation [1]{[#1]}%
	\providecommand \BibitemOpen [0]{}%
	\providecommand \bibitemStop [0]{}%
	\providecommand \bibitemNoStop [0]{.\EOS\space}%
	\providecommand \EOS [0]{\spacefactor3000\relax}%
	\providecommand \BibitemShut  [1]{\csname bibitem#1\endcsname}%
	\let\auto@bib@innerbib\@empty
	\bibitem [{\citenamefont {Scalapino}(2012)}]{ScalapinoRMP2012}%
	\BibitemOpen
	\bibfield  {author} {\bibinfo {author} {\bibfnamefont {D.~J.}\ \bibnamefont
			{Scalapino}},\ }\href {\doibase 10.1103/RevModPhys.84.1383} {\bibfield
		{journal} {\bibinfo  {journal} {Rev. Mod. Phys.}\ }\textbf {\bibinfo {volume}
			{84}},\ \bibinfo {pages} {1383} (\bibinfo {year} {2012})}\BibitemShut
	{NoStop}%
	\bibitem [{\citenamefont {Jang}\ \emph {et~al.}(2016)\citenamefont {Jang},
		\citenamefont {Sakakibara}, \citenamefont {Kino}, \citenamefont {Kotani},
		\citenamefont {Kuroki},\ and\ \citenamefont {Han}}]{JangSciRep2016}%
	\BibitemOpen
	\bibfield  {author} {\bibinfo {author} {\bibfnamefont {S.~W.}\ \bibnamefont
			{Jang}}, \bibinfo {author} {\bibfnamefont {H.}~\bibnamefont {Sakakibara}},
		\bibinfo {author} {\bibfnamefont {H.}~\bibnamefont {Kino}}, \bibinfo {author}
		{\bibfnamefont {T.}~\bibnamefont {Kotani}}, \bibinfo {author} {\bibfnamefont
			{K.}~\bibnamefont {Kuroki}}, \ and\ \bibinfo {author} {\bibfnamefont {M.~J.}\
			\bibnamefont {Han}},\ }\href {\doibase 10.1038/srep33397} {\bibfield
		{journal} {\bibinfo  {journal} {Sci. Rep.}\ }\textbf {\bibinfo {volume}
			{6}},\ \bibinfo {pages} {33397} (\bibinfo {year} {2016})}\BibitemShut
	{NoStop}%
	\bibitem [{\citenamefont {Hirayama}\ \emph {et~al.}(2018)\citenamefont
		{Hirayama}, \citenamefont {Yamaji}, \citenamefont {Misawa},\ and\
		\citenamefont {Imada}}]{HirayamaPRB2018}%
	\BibitemOpen
	\bibfield  {author} {\bibinfo {author} {\bibfnamefont {M.}~\bibnamefont
			{Hirayama}}, \bibinfo {author} {\bibfnamefont {Y.}~\bibnamefont {Yamaji}},
		\bibinfo {author} {\bibfnamefont {T.}~\bibnamefont {Misawa}}, \ and\ \bibinfo
		{author} {\bibfnamefont {M.}~\bibnamefont {Imada}},\ }\href {\doibase
		10.1103/PhysRevB.98.134501} {\bibfield  {journal} {\bibinfo  {journal} {Phys.
				Rev. B}\ }\textbf {\bibinfo {volume} {98}},\ \bibinfo {pages} {134501}
		(\bibinfo {year} {2018})}\BibitemShut {NoStop}%
	\bibitem [{\citenamefont {Misawa}\ and\ \citenamefont
		{Imada}(2014)}]{MisawaPRB2014}%
	\BibitemOpen
	\bibfield  {author} {\bibinfo {author} {\bibfnamefont {T.}~\bibnamefont
			{Misawa}}\ and\ \bibinfo {author} {\bibfnamefont {M.}~\bibnamefont {Imada}},\
	}\href {\doibase 10.1103/PhysRevB.90.115137} {\bibfield  {journal} {\bibinfo
			{journal} {Phys. Rev. B}\ }\textbf {\bibinfo {volume} {90}},\ \bibinfo
		{pages} {115137} (\bibinfo {year} {2014})}\BibitemShut {NoStop}%
	\bibitem [{\citenamefont {Ido}\ \emph {et~al.}(2018)\citenamefont {Ido},
		\citenamefont {Ohgoe},\ and\ \citenamefont {Imada}}]{IdoPRB2018}%
	\BibitemOpen
	\bibfield  {author} {\bibinfo {author} {\bibfnamefont {K.}~\bibnamefont
			{Ido}}, \bibinfo {author} {\bibfnamefont {T.}~\bibnamefont {Ohgoe}}, \ and\
		\bibinfo {author} {\bibfnamefont {M.}~\bibnamefont {Imada}},\ }\href
	{\doibase 10.1103/PhysRevB.97.045138} {\bibfield  {journal} {\bibinfo
			{journal} {Phys. Rev. B}\ }\textbf {\bibinfo {volume} {97}},\ \bibinfo
		{pages} {045138} (\bibinfo {year} {2018})}\BibitemShut {NoStop}%
	\bibitem [{\citenamefont {Imada}(2021)}]{ImadaJPSJ2021}%
	\BibitemOpen
	\bibfield  {author} {\bibinfo {author} {\bibfnamefont {M.}~\bibnamefont
			{Imada}},\ }\href {\doibase 10.7566/JPSJ.90.111009} {\bibfield  {journal}
		{\bibinfo  {journal} {J. Phys. Soc. Jpn.}\ }\textbf {\bibinfo {volume}
			{90}},\ \bibinfo {pages} {111009} (\bibinfo {year} {2021})}\BibitemShut
	{NoStop}%
	\bibitem [{\citenamefont {Momono}\ \emph {et~al.}(1994)\citenamefont {Momono},
		\citenamefont {Ido}, \citenamefont {Nakano}, \citenamefont {Oda},
		\citenamefont {Okajima},\ and\ \citenamefont {Yamaya}}]{MomonoPhysC1994}%
	\BibitemOpen
	\bibfield  {author} {\bibinfo {author} {\bibfnamefont {N.}~\bibnamefont
			{Momono}}, \bibinfo {author} {\bibfnamefont {M.}~\bibnamefont {Ido}},
		\bibinfo {author} {\bibfnamefont {T.}~\bibnamefont {Nakano}}, \bibinfo
		{author} {\bibfnamefont {M.}~\bibnamefont {Oda}}, \bibinfo {author}
		{\bibfnamefont {Y.}~\bibnamefont {Okajima}}, \ and\ \bibinfo {author}
		{\bibfnamefont {K.}~\bibnamefont {Yamaya}},\ }\href {\doibase
		https://doi.org/10.1016/0921-4534(94)90768-4} {\bibfield  {journal} {\bibinfo
			{journal} {Physica C}\ }\textbf {\bibinfo {volume} {233}},\ \bibinfo {pages}
		{395 } (\bibinfo {year} {1994})}\BibitemShut {NoStop}%
	\bibitem [{\citenamefont {Kakeshita}\ \emph {et~al.}(2009)\citenamefont
		{Kakeshita}, \citenamefont {Adachi},\ and\ \citenamefont
		{Uchida}}]{KakeshitaJPC2009}%
	\BibitemOpen
	\bibfield  {author} {\bibinfo {author} {\bibfnamefont {T.}~\bibnamefont
			{Kakeshita}}, \bibinfo {author} {\bibfnamefont {S.}~\bibnamefont {Adachi}}, \
		and\ \bibinfo {author} {\bibfnamefont {S.}~\bibnamefont {Uchida}},\ }\href
	{\doibase 10.1088/1742-6596/150/5/052089} {\bibfield  {journal} {\bibinfo
			{journal} {J. Phys.: Conf. Ser.}\ }\textbf {\bibinfo {volume} {150}},\
		\bibinfo {pages} {052089} (\bibinfo {year} {2009})}\BibitemShut {NoStop}%
	\bibitem [{\citenamefont {Yoshida}\ \emph {et~al.}(2007)\citenamefont
		{Yoshida}, \citenamefont {Zhou}, \citenamefont {Lu}, \citenamefont {Komiya},
		\citenamefont {Ando}, \citenamefont {Eisaki}, \citenamefont {Kakeshita},
		\citenamefont {Uchida}, \citenamefont {Hussain}, \citenamefont {Shen},\ and\
		\citenamefont {Fujimori}}]{YoshidaJCMP07}%
	\BibitemOpen
	\bibfield  {author} {\bibinfo {author} {\bibfnamefont {T.}~\bibnamefont
			{Yoshida}}, \bibinfo {author} {\bibfnamefont {X.~J.}\ \bibnamefont {Zhou}},
		\bibinfo {author} {\bibfnamefont {D.~H.}\ \bibnamefont {Lu}}, \bibinfo
		{author} {\bibfnamefont {S.}~\bibnamefont {Komiya}}, \bibinfo {author}
		{\bibfnamefont {Y.}~\bibnamefont {Ando}}, \bibinfo {author} {\bibfnamefont
			{H.}~\bibnamefont {Eisaki}}, \bibinfo {author} {\bibfnamefont
			{T.}~\bibnamefont {Kakeshita}}, \bibinfo {author} {\bibfnamefont
			{S.}~\bibnamefont {Uchida}}, \bibinfo {author} {\bibfnamefont
			{Z.}~\bibnamefont {Hussain}}, \bibinfo {author} {\bibfnamefont {Z.-X.}\
			\bibnamefont {Shen}}, \ and\ \bibinfo {author} {\bibfnamefont
			{A.}~\bibnamefont {Fujimori}},\ }\href {\doibase
		10.1088/0953-8984/19/12/125209} {\bibfield  {journal} {\bibinfo  {journal}
			{J. Phys. Condens. Matter.}\ }\textbf {\bibinfo {volume} {19}},\ \bibinfo
		{pages} {125209} (\bibinfo {year} {2007})}\BibitemShut {NoStop}%
	\bibitem [{\citenamefont {Armitage}\ \emph {et~al.}(2010)\citenamefont
		{Armitage}, \citenamefont {Fournier},\ and\ \citenamefont
		{Greene}}]{ArmitageRMP2010}%
	\BibitemOpen
	\bibfield  {author} {\bibinfo {author} {\bibfnamefont {N.~P.}\ \bibnamefont
			{Armitage}}, \bibinfo {author} {\bibfnamefont {P.}~\bibnamefont {Fournier}},
		\ and\ \bibinfo {author} {\bibfnamefont {R.~L.}\ \bibnamefont {Greene}},\
	}\href {\doibase 10.1103/RevModPhys.82.2421} {\bibfield  {journal} {\bibinfo
			{journal} {Rev. Mod. Phys.}\ }\textbf {\bibinfo {volume} {82}},\ \bibinfo
		{pages} {2421} (\bibinfo {year} {2010})}\BibitemShut {NoStop}%
	\bibitem [{\citenamefont {Tokura}\ \emph {et~al.}(1990)\citenamefont {Tokura},
		\citenamefont {Koshihara}, \citenamefont {Arima}, \citenamefont {Takagi},
		\citenamefont {Ishibashi}, \citenamefont {Ido},\ and\ \citenamefont
		{Uchida}}]{TokuraPRB1990}%
	\BibitemOpen
	\bibfield  {author} {\bibinfo {author} {\bibfnamefont {Y.}~\bibnamefont
			{Tokura}}, \bibinfo {author} {\bibfnamefont {S.}~\bibnamefont {Koshihara}},
		\bibinfo {author} {\bibfnamefont {T.}~\bibnamefont {Arima}}, \bibinfo
		{author} {\bibfnamefont {H.}~\bibnamefont {Takagi}}, \bibinfo {author}
		{\bibfnamefont {S.}~\bibnamefont {Ishibashi}}, \bibinfo {author}
		{\bibfnamefont {T.}~\bibnamefont {Ido}}, \ and\ \bibinfo {author}
		{\bibfnamefont {S.}~\bibnamefont {Uchida}},\ }\href {\doibase
		10.1103/PhysRevB.41.11657} {\bibfield  {journal} {\bibinfo  {journal} {Phys.
				Rev. B}\ }\textbf {\bibinfo {volume} {41}},\ \bibinfo {pages} {11657}
		(\bibinfo {year} {1990})}\BibitemShut {NoStop}%
	\bibitem [{\citenamefont {Takamatsu}\ \emph {et~al.}(2012)\citenamefont
		{Takamatsu}, \citenamefont {Kato}, \citenamefont {Noji},\ and\ \citenamefont
		{Koike}}]{TakamatsuAPEX2012}%
	\BibitemOpen
	\bibfield  {author} {\bibinfo {author} {\bibfnamefont {T.}~\bibnamefont
			{Takamatsu}}, \bibinfo {author} {\bibfnamefont {M.}~\bibnamefont {Kato}},
		\bibinfo {author} {\bibfnamefont {T.}~\bibnamefont {Noji}}, \ and\ \bibinfo
		{author} {\bibfnamefont {Y.}~\bibnamefont {Koike}},\ }\href {\doibase
		10.1143/APEX.5.073101} {\bibfield  {journal} {\bibinfo  {journal} {Appl.
				Phys. Express}\ }\textbf {\bibinfo {volume} {5}},\ \bibinfo {pages} {073101}
		(\bibinfo {year} {2012})}\BibitemShut {NoStop}%
	\bibitem [{\citenamefont {Lin}\ \emph {et~al.}(2019)\citenamefont {Lin},
		\citenamefont {Horio}, \citenamefont {Kawamata}, \citenamefont {Saito},
		\citenamefont {Koshiishi}, \citenamefont {Sakamoto}, \citenamefont {Zhang},
		\citenamefont {Yamamoto}, \citenamefont {Ikeda}, \citenamefont {Hirata},
		\citenamefont {Takubo}, \citenamefont {Wadati}, \citenamefont {Yasui},
		\citenamefont {Takagi}, \citenamefont {Ikenaga}, \citenamefont {Adachi},
		\citenamefont {Koike},\ and\ \citenamefont {Fujimori}}]{LinJPSJ2019}%
	\BibitemOpen
	\bibfield  {author} {\bibinfo {author} {\bibfnamefont {C.}~\bibnamefont
			{Lin}}, \bibinfo {author} {\bibfnamefont {M.}~\bibnamefont {Horio}}, \bibinfo
		{author} {\bibfnamefont {T.}~\bibnamefont {Kawamata}}, \bibinfo {author}
		{\bibfnamefont {S.}~\bibnamefont {Saito}}, \bibinfo {author} {\bibfnamefont
			{K.}~\bibnamefont {Koshiishi}}, \bibinfo {author} {\bibfnamefont
			{S.}~\bibnamefont {Sakamoto}}, \bibinfo {author} {\bibfnamefont
			{Y.}~\bibnamefont {Zhang}}, \bibinfo {author} {\bibfnamefont
			{K.}~\bibnamefont {Yamamoto}}, \bibinfo {author} {\bibfnamefont
			{K.}~\bibnamefont {Ikeda}}, \bibinfo {author} {\bibfnamefont
			{Y.}~\bibnamefont {Hirata}}, \bibinfo {author} {\bibfnamefont
			{K.}~\bibnamefont {Takubo}}, \bibinfo {author} {\bibfnamefont
			{H.}~\bibnamefont {Wadati}}, \bibinfo {author} {\bibfnamefont
			{A.}~\bibnamefont {Yasui}}, \bibinfo {author} {\bibfnamefont
			{Y.}~\bibnamefont {Takagi}}, \bibinfo {author} {\bibfnamefont
			{E.}~\bibnamefont {Ikenaga}}, \bibinfo {author} {\bibfnamefont
			{T.}~\bibnamefont {Adachi}}, \bibinfo {author} {\bibfnamefont
			{Y.}~\bibnamefont {Koike}}, \ and\ \bibinfo {author} {\bibfnamefont
			{A.}~\bibnamefont {Fujimori}},\ }\href {\doibase 10.7566/JPSJ.88.115004}
	{\bibfield  {journal} {\bibinfo  {journal} {J. Phys. Soc. Jpn.}\ }\textbf
		{\bibinfo {volume} {88}},\ \bibinfo {pages} {115004} (\bibinfo {year}
		{2019})}\BibitemShut {NoStop}%
	\bibitem [{\citenamefont {Akimitsu}\ \emph {et~al.}(1988)\citenamefont
		{Akimitsu}, \citenamefont {Suzuki}, \citenamefont {Watanabe},\ and\
		\citenamefont {Sawa}}]{AkimitsuJJAP1988}%
	\BibitemOpen
	\bibfield  {author} {\bibinfo {author} {\bibfnamefont {J.}~\bibnamefont
			{Akimitsu}}, \bibinfo {author} {\bibfnamefont {S.}~\bibnamefont {Suzuki}},
		\bibinfo {author} {\bibfnamefont {M.}~\bibnamefont {Watanabe}}, \ and\
		\bibinfo {author} {\bibfnamefont {H.}~\bibnamefont {Sawa}},\ }\href {\doibase
		10.1143/JJAP.27.L1859} {\bibfield  {journal} {\bibinfo  {journal} {Jpn. J.
				Appl. Phys.}\ }\textbf {\bibinfo {volume} {27}},\ \bibinfo {pages} {L1859}
		(\bibinfo {year} {1988})}\BibitemShut {NoStop}%
	\bibitem [{\citenamefont {Izumi}\ \emph {et~al.}(1989)\citenamefont {Izumi},
		\citenamefont {Takayama-Muromachi}, \citenamefont {Fujimori}, \citenamefont
		{Kamiyama}, \citenamefont {Asano}, \citenamefont {Akimitsu},\ and\
		\citenamefont {Sawa}}]{IzumiPhysC1989}%
	\BibitemOpen
	\bibfield  {author} {\bibinfo {author} {\bibfnamefont {F.}~\bibnamefont
			{Izumi}}, \bibinfo {author} {\bibfnamefont {E.}~\bibnamefont
			{Takayama-Muromachi}}, \bibinfo {author} {\bibfnamefont {A.}~\bibnamefont
			{Fujimori}}, \bibinfo {author} {\bibfnamefont {T.}~\bibnamefont {Kamiyama}},
		\bibinfo {author} {\bibfnamefont {H.}~\bibnamefont {Asano}}, \bibinfo
		{author} {\bibfnamefont {J.}~\bibnamefont {Akimitsu}}, \ and\ \bibinfo
		{author} {\bibfnamefont {H.}~\bibnamefont {Sawa}},\ }\href {\doibase
		https://doi.org/10.1016/0921-4534(89)90241-4} {\bibfield  {journal} {\bibinfo
			{journal} {Physica C}\ }\textbf {\bibinfo {volume} {158}},\ \bibinfo {pages}
		{440} (\bibinfo {year} {1989})}\BibitemShut {NoStop}%
	\bibitem [{\citenamefont {Asano}\ \emph {et~al.}(2019)\citenamefont {Asano},
		\citenamefont {Suzuki}, \citenamefont {Kudo}, \citenamefont {Watanabe},
		\citenamefont {Koda}, \citenamefont {Kadono}, \citenamefont {Noji},
		\citenamefont {Koike}, \citenamefont {Taniguchi}, \citenamefont {Kitagawa},
		\citenamefont {Ishida},\ and\ \citenamefont {Fujita}}]{AsanoJPSJ2019}%
	\BibitemOpen
	\bibfield  {author} {\bibinfo {author} {\bibfnamefont {S.}~\bibnamefont
			{Asano}}, \bibinfo {author} {\bibfnamefont {K.~M.}\ \bibnamefont {Suzuki}},
		\bibinfo {author} {\bibfnamefont {K.}~\bibnamefont {Kudo}}, \bibinfo {author}
		{\bibfnamefont {I.}~\bibnamefont {Watanabe}}, \bibinfo {author}
		{\bibfnamefont {A.}~\bibnamefont {Koda}}, \bibinfo {author} {\bibfnamefont
			{R.}~\bibnamefont {Kadono}}, \bibinfo {author} {\bibfnamefont
			{T.}~\bibnamefont {Noji}}, \bibinfo {author} {\bibfnamefont {Y.}~\bibnamefont
			{Koike}}, \bibinfo {author} {\bibfnamefont {T.}~\bibnamefont {Taniguchi}},
		\bibinfo {author} {\bibfnamefont {S.}~\bibnamefont {Kitagawa}}, \bibinfo
		{author} {\bibfnamefont {K.}~\bibnamefont {Ishida}}, \ and\ \bibinfo {author}
		{\bibfnamefont {M.}~\bibnamefont {Fujita}},\ }\href {\doibase
		10.7566/JPSJ.88.084709} {\bibfield  {journal} {\bibinfo  {journal} {J. Phys.
				Soc. Jpn.}\ }\textbf {\bibinfo {volume} {88}},\ \bibinfo {pages} {084709}
		(\bibinfo {year} {2019})}\BibitemShut {NoStop}%
	\bibitem [{\citenamefont {Asano}\ \emph {et~al.}(2020)\citenamefont {Asano},
		\citenamefont {Ishii}, \citenamefont {Yamagami}, \citenamefont {Miyawaki},
		\citenamefont {Harada},\ and\ \citenamefont {Fujita}}]{AsanoJPSJ2020}%
	\BibitemOpen
	\bibfield  {author} {\bibinfo {author} {\bibfnamefont {S.}~\bibnamefont
			{Asano}}, \bibinfo {author} {\bibfnamefont {K.}~\bibnamefont {Ishii}},
		\bibinfo {author} {\bibfnamefont {K.}~\bibnamefont {Yamagami}}, \bibinfo
		{author} {\bibfnamefont {J.}~\bibnamefont {Miyawaki}}, \bibinfo {author}
		{\bibfnamefont {Y.}~\bibnamefont {Harada}}, \ and\ \bibinfo {author}
		{\bibfnamefont {M.}~\bibnamefont {Fujita}},\ }\href {\doibase
		10.7566/JPSJ.89.075002} {\bibfield  {journal} {\bibinfo  {journal} {J. Phys.
				Soc. Jpn.}\ }\textbf {\bibinfo {volume} {89}},\ \bibinfo {pages} {075002}
		(\bibinfo {year} {2020})}\BibitemShut {NoStop}%
	\bibitem [{\citenamefont {Ino}\ \emph {et~al.}(2004)\citenamefont {Ino},
		\citenamefont {Higashiguchi}, \citenamefont {Yamazaki}, \citenamefont
		{Yamasaki}, \citenamefont {Narimura}, \citenamefont {Kobayashi},
		\citenamefont {Shimada}, \citenamefont {Namatame}, \citenamefont {Taniguchi},
		\citenamefont {Yoshida}, \citenamefont {Fujimori}, \citenamefont {Shen},
		\citenamefont {Kakeshita}, \citenamefont {Uchida}, \citenamefont {Adachi},\
		and\ \citenamefont {Tajima}}]{InoPhysB2004}%
	\BibitemOpen
	\bibfield  {author} {\bibinfo {author} {\bibfnamefont {A.}~\bibnamefont
			{Ino}}, \bibinfo {author} {\bibfnamefont {M.}~\bibnamefont {Higashiguchi}},
		\bibinfo {author} {\bibfnamefont {K.}~\bibnamefont {Yamazaki}}, \bibinfo
		{author} {\bibfnamefont {T.}~\bibnamefont {Yamasaki}}, \bibinfo {author}
		{\bibfnamefont {T.}~\bibnamefont {Narimura}}, \bibinfo {author}
		{\bibfnamefont {K.}~\bibnamefont {Kobayashi}}, \bibinfo {author}
		{\bibfnamefont {K.}~\bibnamefont {Shimada}}, \bibinfo {author} {\bibfnamefont
			{H.}~\bibnamefont {Namatame}}, \bibinfo {author} {\bibfnamefont
			{M.}~\bibnamefont {Taniguchi}}, \bibinfo {author} {\bibfnamefont
			{T.}~\bibnamefont {Yoshida}}, \bibinfo {author} {\bibfnamefont
			{A.}~\bibnamefont {Fujimori}}, \bibinfo {author} {\bibfnamefont {Z.-X.}\
			\bibnamefont {Shen}}, \bibinfo {author} {\bibfnamefont {T.}~\bibnamefont
			{Kakeshita}}, \bibinfo {author} {\bibfnamefont {S.}~\bibnamefont {Uchida}},
		\bibinfo {author} {\bibfnamefont {S.}~\bibnamefont {Adachi}}, \ and\ \bibinfo
		{author} {\bibfnamefont {S.}~\bibnamefont {Tajima}},\ }\href {\doibase
		https://doi.org/10.1016/j.physb.2004.06.024} {\bibfield  {journal} {\bibinfo
			{journal} {Physica B}\ }\textbf {\bibinfo {volume} {351}},\ \bibinfo {pages}
		{274} (\bibinfo {year} {2004})}\BibitemShut {NoStop}%
	\bibitem [{\citenamefont {Yoshida}\ \emph {et~al.}(2003)\citenamefont
		{Yoshida}, \citenamefont {Zhou}, \citenamefont {Sasagawa}, \citenamefont
		{Yang}, \citenamefont {Bogdanov}, \citenamefont {Lanzara}, \citenamefont
		{Hussain}, \citenamefont {Mizokawa}, \citenamefont {Fujimori}, \citenamefont
		{Eisaki}, \citenamefont {Shen}, \citenamefont {Kakeshita},\ and\
		\citenamefont {Uchida}}]{YoshidaPRL2003}%
	\BibitemOpen
	\bibfield  {author} {\bibinfo {author} {\bibfnamefont {T.}~\bibnamefont
			{Yoshida}}, \bibinfo {author} {\bibfnamefont {X.~J.}\ \bibnamefont {Zhou}},
		\bibinfo {author} {\bibfnamefont {T.}~\bibnamefont {Sasagawa}}, \bibinfo
		{author} {\bibfnamefont {W.~L.}\ \bibnamefont {Yang}}, \bibinfo {author}
		{\bibfnamefont {P.~V.}\ \bibnamefont {Bogdanov}}, \bibinfo {author}
		{\bibfnamefont {A.}~\bibnamefont {Lanzara}}, \bibinfo {author} {\bibfnamefont
			{Z.}~\bibnamefont {Hussain}}, \bibinfo {author} {\bibfnamefont
			{T.}~\bibnamefont {Mizokawa}}, \bibinfo {author} {\bibfnamefont
			{A.}~\bibnamefont {Fujimori}}, \bibinfo {author} {\bibfnamefont
			{H.}~\bibnamefont {Eisaki}}, \bibinfo {author} {\bibfnamefont {Z.-X.}\
			\bibnamefont {Shen}}, \bibinfo {author} {\bibfnamefont {T.}~\bibnamefont
			{Kakeshita}}, \ and\ \bibinfo {author} {\bibfnamefont {S.}~\bibnamefont
			{Uchida}},\ }\href {\doibase 10.1103/PhysRevLett.91.027001} {\bibfield
		{journal} {\bibinfo  {journal} {Phys. Rev. Lett.}\ }\textbf {\bibinfo
			{volume} {91}},\ \bibinfo {pages} {027001} (\bibinfo {year}
		{2003})}\BibitemShut {NoStop}%
	\bibitem [{\citenamefont {Shen}\ \emph {et~al.}(2004)\citenamefont {Shen},
		\citenamefont {Ronning}, \citenamefont {Lu}, \citenamefont {Lee},
		\citenamefont {Ingle}, \citenamefont {Meevasana}, \citenamefont {Baumberger},
		\citenamefont {Damascelli}, \citenamefont {Armitage}, \citenamefont {Miller},
		\citenamefont {Kohsaka}, \citenamefont {Azuma}, \citenamefont {Takano},
		\citenamefont {Takagi},\ and\ \citenamefont {Shen}}]{ShenPRL2004}%
	\BibitemOpen
	\bibfield  {author} {\bibinfo {author} {\bibfnamefont {K.~M.}\ \bibnamefont
			{Shen}}, \bibinfo {author} {\bibfnamefont {F.}~\bibnamefont {Ronning}},
		\bibinfo {author} {\bibfnamefont {D.~H.}\ \bibnamefont {Lu}}, \bibinfo
		{author} {\bibfnamefont {W.~S.}\ \bibnamefont {Lee}}, \bibinfo {author}
		{\bibfnamefont {N.~J.~C.}\ \bibnamefont {Ingle}}, \bibinfo {author}
		{\bibfnamefont {W.}~\bibnamefont {Meevasana}}, \bibinfo {author}
		{\bibfnamefont {F.}~\bibnamefont {Baumberger}}, \bibinfo {author}
		{\bibfnamefont {A.}~\bibnamefont {Damascelli}}, \bibinfo {author}
		{\bibfnamefont {N.~P.}\ \bibnamefont {Armitage}}, \bibinfo {author}
		{\bibfnamefont {L.~L.}\ \bibnamefont {Miller}}, \bibinfo {author}
		{\bibfnamefont {Y.}~\bibnamefont {Kohsaka}}, \bibinfo {author} {\bibfnamefont
			{M.}~\bibnamefont {Azuma}}, \bibinfo {author} {\bibfnamefont
			{M.}~\bibnamefont {Takano}}, \bibinfo {author} {\bibfnamefont
			{H.}~\bibnamefont {Takagi}}, \ and\ \bibinfo {author} {\bibfnamefont {Z.-X.}\
			\bibnamefont {Shen}},\ }\href {\doibase 10.1103/PhysRevLett.93.267002}
	{\bibfield  {journal} {\bibinfo  {journal} {Phys. Rev. Lett.}\ }\textbf
		{\bibinfo {volume} {93}},\ \bibinfo {pages} {267002} (\bibinfo {year}
		{2004})}\BibitemShut {NoStop}%
	\bibitem [{\citenamefont {Kohsaka}\ \emph {et~al.}(2003)\citenamefont
		{Kohsaka}, \citenamefont {Sasagawa}, \citenamefont {Ronning}, \citenamefont
		{Yoshida}, \citenamefont {Kim}, \citenamefont {Hanaguri}, \citenamefont
		{Azuma}, \citenamefont {Takano}, \citenamefont {Xun~Shen},\ and\
		\citenamefont {Takagi}}]{KohsakaJPSJ2003}%
	\BibitemOpen
	\bibfield  {author} {\bibinfo {author} {\bibfnamefont {Y.}~\bibnamefont
			{Kohsaka}}, \bibinfo {author} {\bibfnamefont {T.}~\bibnamefont {Sasagawa}},
		\bibinfo {author} {\bibfnamefont {F.}~\bibnamefont {Ronning}}, \bibinfo
		{author} {\bibfnamefont {T.}~\bibnamefont {Yoshida}}, \bibinfo {author}
		{\bibfnamefont {C.}~\bibnamefont {Kim}}, \bibinfo {author} {\bibfnamefont
			{T.}~\bibnamefont {Hanaguri}}, \bibinfo {author} {\bibfnamefont
			{M.}~\bibnamefont {Azuma}}, \bibinfo {author} {\bibfnamefont
			{M.}~\bibnamefont {Takano}}, \bibinfo {author} {\bibfnamefont
			{Z.}~\bibnamefont {Xun~Shen}}, \ and\ \bibinfo {author} {\bibfnamefont
			{H.}~\bibnamefont {Takagi}},\ }\href {\doibase 10.1143/JPSJ.72.1018}
	{\bibfield  {journal} {\bibinfo  {journal} {J. Phys. Soc. Jpn.}\ }\textbf
		{\bibinfo {volume} {72}},\ \bibinfo {pages} {1018} (\bibinfo {year}
		{2003})}\BibitemShut {NoStop}%
	\bibitem [{\citenamefont {Hashimoto}\ \emph {et~al.}(2008)\citenamefont
		{Hashimoto}, \citenamefont {Yoshida}, \citenamefont {Yagi}, \citenamefont
		{Takizawa}, \citenamefont {Fujimori}, \citenamefont {Kubota}, \citenamefont
		{Ono}, \citenamefont {Tanaka}, \citenamefont {Lu}, \citenamefont {Shen},
		\citenamefont {Ono},\ and\ \citenamefont {Ando}}]{HashimotoPRB2008}%
	\BibitemOpen
	\bibfield  {author} {\bibinfo {author} {\bibfnamefont {M.}~\bibnamefont
			{Hashimoto}}, \bibinfo {author} {\bibfnamefont {T.}~\bibnamefont {Yoshida}},
		\bibinfo {author} {\bibfnamefont {H.}~\bibnamefont {Yagi}}, \bibinfo {author}
		{\bibfnamefont {M.}~\bibnamefont {Takizawa}}, \bibinfo {author}
		{\bibfnamefont {A.}~\bibnamefont {Fujimori}}, \bibinfo {author}
		{\bibfnamefont {M.}~\bibnamefont {Kubota}}, \bibinfo {author} {\bibfnamefont
			{K.}~\bibnamefont {Ono}}, \bibinfo {author} {\bibfnamefont {K.}~\bibnamefont
			{Tanaka}}, \bibinfo {author} {\bibfnamefont {D.~H.}\ \bibnamefont {Lu}},
		\bibinfo {author} {\bibfnamefont {Z.-X.}\ \bibnamefont {Shen}}, \bibinfo
		{author} {\bibfnamefont {S.}~\bibnamefont {Ono}}, \ and\ \bibinfo {author}
		{\bibfnamefont {Y.}~\bibnamefont {Ando}},\ }\href {\doibase
		10.1103/PhysRevB.77.094516} {\bibfield  {journal} {\bibinfo  {journal} {Phys.
				Rev. B}\ }\textbf {\bibinfo {volume} {77}},\ \bibinfo {pages} {094516}
		(\bibinfo {year} {2008})}\BibitemShut {NoStop}%
	\bibitem [{\citenamefont {Peng}\ \emph {et~al.}(2013)\citenamefont {Peng},
		\citenamefont {Meng}, \citenamefont {Mou}, \citenamefont {He}, \citenamefont
		{Zhao}, \citenamefont {Wu}, \citenamefont {Liu}, \citenamefont {Dong},
		\citenamefont {He}, \citenamefont {Zhang}, \citenamefont {Wang},
		\citenamefont {Peng}, \citenamefont {Wang}, \citenamefont {Zhang},
		\citenamefont {Yang}, \citenamefont {Chen}, \citenamefont {Xu}, \citenamefont
		{Lee},\ and\ \citenamefont {Zhou}}]{PengNatCommun2013}%
	\BibitemOpen
	\bibfield  {author} {\bibinfo {author} {\bibfnamefont {Y.}~\bibnamefont
			{Peng}}, \bibinfo {author} {\bibfnamefont {J.}~\bibnamefont {Meng}}, \bibinfo
		{author} {\bibfnamefont {D.}~\bibnamefont {Mou}}, \bibinfo {author}
		{\bibfnamefont {J.}~\bibnamefont {He}}, \bibinfo {author} {\bibfnamefont
			{L.}~\bibnamefont {Zhao}}, \bibinfo {author} {\bibfnamefont {Y.}~\bibnamefont
			{Wu}}, \bibinfo {author} {\bibfnamefont {G.}~\bibnamefont {Liu}}, \bibinfo
		{author} {\bibfnamefont {X.}~\bibnamefont {Dong}}, \bibinfo {author}
		{\bibfnamefont {S.}~\bibnamefont {He}}, \bibinfo {author} {\bibfnamefont
			{J.}~\bibnamefont {Zhang}}, \bibinfo {author} {\bibfnamefont
			{X.}~\bibnamefont {Wang}}, \bibinfo {author} {\bibfnamefont {Q.}~\bibnamefont
			{Peng}}, \bibinfo {author} {\bibfnamefont {Z.}~\bibnamefont {Wang}}, \bibinfo
		{author} {\bibfnamefont {S.}~\bibnamefont {Zhang}}, \bibinfo {author}
		{\bibfnamefont {F.}~\bibnamefont {Yang}}, \bibinfo {author} {\bibfnamefont
			{C.}~\bibnamefont {Chen}}, \bibinfo {author} {\bibfnamefont {Z.}~\bibnamefont
			{Xu}}, \bibinfo {author} {\bibfnamefont {T.~K.}\ \bibnamefont {Lee}}, \ and\
		\bibinfo {author} {\bibfnamefont {X.~J.}\ \bibnamefont {Zhou}},\ }\href
	{\doibase 10.1038/ncomms3459} {\bibfield  {journal} {\bibinfo  {journal}
			{Nat. Commun.}\ }\textbf {\bibinfo {volume} {4}},\ \bibinfo {pages} {2459}
		(\bibinfo {year} {2013})}\BibitemShut {NoStop}%
	\bibitem [{\citenamefont {Tanaka}\ \emph {et~al.}(2010)\citenamefont {Tanaka},
		\citenamefont {Yoshida}, \citenamefont {Shen}, \citenamefont {Lu},
		\citenamefont {Lee}, \citenamefont {Yagi}, \citenamefont {Fujimori},
		\citenamefont {Shen}, \citenamefont {Risdiana}, \citenamefont {Fujii},\ and\
		\citenamefont {Terasaki}}]{TanakaPRB2010}%
	\BibitemOpen
	\bibfield  {author} {\bibinfo {author} {\bibfnamefont {K.}~\bibnamefont
			{Tanaka}}, \bibinfo {author} {\bibfnamefont {T.}~\bibnamefont {Yoshida}},
		\bibinfo {author} {\bibfnamefont {K.~M.}\ \bibnamefont {Shen}}, \bibinfo
		{author} {\bibfnamefont {D.~H.}\ \bibnamefont {Lu}}, \bibinfo {author}
		{\bibfnamefont {W.~S.}\ \bibnamefont {Lee}}, \bibinfo {author} {\bibfnamefont
			{H.}~\bibnamefont {Yagi}}, \bibinfo {author} {\bibfnamefont {A.}~\bibnamefont
			{Fujimori}}, \bibinfo {author} {\bibfnamefont {Z.-X.}\ \bibnamefont {Shen}},
		\bibinfo {author} {\bibnamefont {Risdiana}}, \bibinfo {author} {\bibfnamefont
			{T.}~\bibnamefont {Fujii}}, \ and\ \bibinfo {author} {\bibfnamefont
			{I.}~\bibnamefont {Terasaki}},\ }\href {\doibase 10.1103/PhysRevB.81.125115}
	{\bibfield  {journal} {\bibinfo  {journal} {Phys. Rev. B}\ }\textbf {\bibinfo
			{volume} {81}},\ \bibinfo {pages} {125115} (\bibinfo {year}
		{2010})}\BibitemShut {NoStop}%
	\bibitem [{\citenamefont {Yoshida}\ \emph {et~al.}(2006)\citenamefont
		{Yoshida}, \citenamefont {Zhou}, \citenamefont {Tanaka}, \citenamefont
		{Yang}, \citenamefont {Hussain}, \citenamefont {Shen}, \citenamefont
		{Fujimori}, \citenamefont {Sahrakorpi}, \citenamefont {Lindroos},
		\citenamefont {Markiewicz}, \citenamefont {Bansil}, \citenamefont {Komiya},
		\citenamefont {Ando}, \citenamefont {Eisaki}, \citenamefont {Kakeshita},\
		and\ \citenamefont {Uchida}}]{YoshidaPRB2006}%
	\BibitemOpen
	\bibfield  {author} {\bibinfo {author} {\bibfnamefont {T.}~\bibnamefont
			{Yoshida}}, \bibinfo {author} {\bibfnamefont {X.~J.}\ \bibnamefont {Zhou}},
		\bibinfo {author} {\bibfnamefont {K.}~\bibnamefont {Tanaka}}, \bibinfo
		{author} {\bibfnamefont {W.~L.}\ \bibnamefont {Yang}}, \bibinfo {author}
		{\bibfnamefont {Z.}~\bibnamefont {Hussain}}, \bibinfo {author} {\bibfnamefont
			{Z.-X.}\ \bibnamefont {Shen}}, \bibinfo {author} {\bibfnamefont
			{A.}~\bibnamefont {Fujimori}}, \bibinfo {author} {\bibfnamefont
			{S.}~\bibnamefont {Sahrakorpi}}, \bibinfo {author} {\bibfnamefont
			{M.}~\bibnamefont {Lindroos}}, \bibinfo {author} {\bibfnamefont {R.~S.}\
			\bibnamefont {Markiewicz}}, \bibinfo {author} {\bibfnamefont
			{A.}~\bibnamefont {Bansil}}, \bibinfo {author} {\bibfnamefont
			{S.}~\bibnamefont {Komiya}}, \bibinfo {author} {\bibfnamefont
			{Y.}~\bibnamefont {Ando}}, \bibinfo {author} {\bibfnamefont {H.}~\bibnamefont
			{Eisaki}}, \bibinfo {author} {\bibfnamefont {T.}~\bibnamefont {Kakeshita}}, \
		and\ \bibinfo {author} {\bibfnamefont {S.}~\bibnamefont {Uchida}},\ }\href
	{\doibase 10.1103/PhysRevB.74.224510} {\bibfield  {journal} {\bibinfo
			{journal} {Phys. Rev. B}\ }\textbf {\bibinfo {volume} {74}},\ \bibinfo
		{pages} {224510} (\bibinfo {year} {2006})}\BibitemShut {NoStop}%
	\bibitem [{\citenamefont {Shen}\ \emph {et~al.}(2005)\citenamefont {Shen},
		\citenamefont {Ronning}, \citenamefont {Lu}, \citenamefont {Baumberger},
		\citenamefont {Ingle}, \citenamefont {Lee}, \citenamefont {Meevasana},
		\citenamefont {Kohsaka}, \citenamefont {Azuma}, \citenamefont {Takano},
		\citenamefont {Takagi},\ and\ \citenamefont {Shen}}]{ShenScience2005}%
	\BibitemOpen
	\bibfield  {author} {\bibinfo {author} {\bibfnamefont {K.~M.}\ \bibnamefont
			{Shen}}, \bibinfo {author} {\bibfnamefont {F.}~\bibnamefont {Ronning}},
		\bibinfo {author} {\bibfnamefont {D.~H.}\ \bibnamefont {Lu}}, \bibinfo
		{author} {\bibfnamefont {F.}~\bibnamefont {Baumberger}}, \bibinfo {author}
		{\bibfnamefont {N.~J.~C.}\ \bibnamefont {Ingle}}, \bibinfo {author}
		{\bibfnamefont {W.~S.}\ \bibnamefont {Lee}}, \bibinfo {author} {\bibfnamefont
			{W.}~\bibnamefont {Meevasana}}, \bibinfo {author} {\bibfnamefont
			{Y.}~\bibnamefont {Kohsaka}}, \bibinfo {author} {\bibfnamefont
			{M.}~\bibnamefont {Azuma}}, \bibinfo {author} {\bibfnamefont
			{M.}~\bibnamefont {Takano}}, \bibinfo {author} {\bibfnamefont
			{H.}~\bibnamefont {Takagi}}, \ and\ \bibinfo {author} {\bibfnamefont {Z.-X.}\
			\bibnamefont {Shen}},\ }\href {\doibase 10.1126/science.1103627} {\bibfield
		{journal} {\bibinfo  {journal} {Science}\ }\textbf {\bibinfo {volume}
			{307}},\ \bibinfo {pages} {901} (\bibinfo {year} {2005})}\BibitemShut
	{NoStop}%
	\bibitem [{\citenamefont {Yang}\ \emph {et~al.}(2006)\citenamefont {Yang},
		\citenamefont {Rice},\ and\ \citenamefont {Zhang}}]{YangPRB2006}%
	\BibitemOpen
	\bibfield  {author} {\bibinfo {author} {\bibfnamefont {K.-Y.}\ \bibnamefont
			{Yang}}, \bibinfo {author} {\bibfnamefont {T.~M.}\ \bibnamefont {Rice}}, \
		and\ \bibinfo {author} {\bibfnamefont {F.-C.}\ \bibnamefont {Zhang}},\ }\href
	{\doibase 10.1103/PhysRevB.73.174501} {\bibfield  {journal} {\bibinfo
			{journal} {Phys. Rev. B}\ }\textbf {\bibinfo {volume} {73}},\ \bibinfo
		{pages} {174501} (\bibinfo {year} {2006})}\BibitemShut {NoStop}%
	\bibitem [{\citenamefont {Yang}\ \emph {et~al.}(2011)\citenamefont {Yang},
		\citenamefont {Rameau}, \citenamefont {Pan}, \citenamefont {Gu},
		\citenamefont {Johnson}, \citenamefont {Claus}, \citenamefont {Hinks},\ and\
		\citenamefont {Kidd}}]{YangPRL2011}%
	\BibitemOpen
	\bibfield  {author} {\bibinfo {author} {\bibfnamefont {H.-B.}\ \bibnamefont
			{Yang}}, \bibinfo {author} {\bibfnamefont {J.~D.}\ \bibnamefont {Rameau}},
		\bibinfo {author} {\bibfnamefont {Z.-H.}\ \bibnamefont {Pan}}, \bibinfo
		{author} {\bibfnamefont {G.~D.}\ \bibnamefont {Gu}}, \bibinfo {author}
		{\bibfnamefont {P.~D.}\ \bibnamefont {Johnson}}, \bibinfo {author}
		{\bibfnamefont {H.}~\bibnamefont {Claus}}, \bibinfo {author} {\bibfnamefont
			{D.~G.}\ \bibnamefont {Hinks}}, \ and\ \bibinfo {author} {\bibfnamefont
			{T.~E.}\ \bibnamefont {Kidd}},\ }\href {\doibase
		10.1103/PhysRevLett.107.047003} {\bibfield  {journal} {\bibinfo  {journal}
			{Phys. Rev. Lett.}\ }\textbf {\bibinfo {volume} {107}},\ \bibinfo {pages}
		{047003} (\bibinfo {year} {2011})}\BibitemShut {NoStop}%
	\bibitem [{\citenamefont {Meng}\ \emph {et~al.}(2011)\citenamefont {Meng},
		\citenamefont {Brunner}, \citenamefont {Kim}, \citenamefont {Lee},
		\citenamefont {Lee}, \citenamefont {Wen}, \citenamefont {Xu}, \citenamefont
		{Gu},\ and\ \citenamefont {Gweon}}]{MengPRB2011}%
	\BibitemOpen
	\bibfield  {author} {\bibinfo {author} {\bibfnamefont {J.-Q.}\ \bibnamefont
			{Meng}}, \bibinfo {author} {\bibfnamefont {M.}~\bibnamefont {Brunner}},
		\bibinfo {author} {\bibfnamefont {K.-H.}\ \bibnamefont {Kim}}, \bibinfo
		{author} {\bibfnamefont {H.-G.}\ \bibnamefont {Lee}}, \bibinfo {author}
		{\bibfnamefont {S.-I.}\ \bibnamefont {Lee}}, \bibinfo {author} {\bibfnamefont
			{J.~S.}\ \bibnamefont {Wen}}, \bibinfo {author} {\bibfnamefont {Z.~J.}\
			\bibnamefont {Xu}}, \bibinfo {author} {\bibfnamefont {G.~D.}\ \bibnamefont
			{Gu}}, \ and\ \bibinfo {author} {\bibfnamefont {G.-H.}\ \bibnamefont
			{Gweon}},\ }\href {\doibase 10.1103/PhysRevB.84.060513} {\bibfield  {journal}
		{\bibinfo  {journal} {Phys. Rev. B}\ }\textbf {\bibinfo {volume} {84}},\
		\bibinfo {pages} {060513} (\bibinfo {year} {2011})}\BibitemShut {NoStop}%
	\bibitem [{\citenamefont {Badoux}\ \emph {et~al.}(2016)\citenamefont {Badoux},
		\citenamefont {Tabis}, \citenamefont {Lalibert\'{e}}, \citenamefont
		{Grissonnanche}, \citenamefont {Vignolle}, \citenamefont {Vignolles},
		\citenamefont {B\'{e}ard}, \citenamefont {Bonn}, \citenamefont {Hardy},
		\citenamefont {Liang}, \citenamefont {Doiron-Leyraud}, \citenamefont
		{Taillefer},\ and\ \citenamefont {Proust}}]{BadouxNature2016}%
	\BibitemOpen
	\bibfield  {author} {\bibinfo {author} {\bibfnamefont {S.}~\bibnamefont
			{Badoux}}, \bibinfo {author} {\bibfnamefont {W.}~\bibnamefont {Tabis}},
		\bibinfo {author} {\bibfnamefont {F.}~\bibnamefont {Lalibert\'{e}}}, \bibinfo
		{author} {\bibfnamefont {G.}~\bibnamefont {Grissonnanche}}, \bibinfo {author}
		{\bibfnamefont {B.}~\bibnamefont {Vignolle}}, \bibinfo {author}
		{\bibfnamefont {D.}~\bibnamefont {Vignolles}}, \bibinfo {author}
		{\bibfnamefont {J.}~\bibnamefont {B\'{e}ard}}, \bibinfo {author}
		{\bibfnamefont {D.~A.}\ \bibnamefont {Bonn}}, \bibinfo {author}
		{\bibfnamefont {W.~N.}\ \bibnamefont {Hardy}}, \bibinfo {author}
		{\bibfnamefont {R.}~\bibnamefont {Liang}}, \bibinfo {author} {\bibfnamefont
			{N.}~\bibnamefont {Doiron-Leyraud}}, \bibinfo {author} {\bibfnamefont
			{L.}~\bibnamefont {Taillefer}}, \ and\ \bibinfo {author} {\bibfnamefont
			{C.}~\bibnamefont {Proust}},\ }\href {\doibase 10.1038/nature16983}
	{\bibfield  {journal} {\bibinfo  {journal} {Nature}\ }\textbf {\bibinfo
			{volume} {531}},\ \bibinfo {pages} {210} (\bibinfo {year}
		{2016})}\BibitemShut {NoStop}%
	\bibitem [{\citenamefont {Collignon}\ \emph {et~al.}(2017)\citenamefont
		{Collignon}, \citenamefont {Badoux}, \citenamefont {Afshar}, \citenamefont
		{Michon}, \citenamefont {Lalibert\'e}, \citenamefont {Cyr-Choini\`ere},
		\citenamefont {Zhou}, \citenamefont {Licciardello}, \citenamefont {Wiedmann},
		\citenamefont {Doiron-Leyraud},\ and\ \citenamefont
		{Taillefer}}]{CollignonPRB2017}%
	\BibitemOpen
	\bibfield  {author} {\bibinfo {author} {\bibfnamefont {C.}~\bibnamefont
			{Collignon}}, \bibinfo {author} {\bibfnamefont {S.}~\bibnamefont {Badoux}},
		\bibinfo {author} {\bibfnamefont {S.~A.~A.}\ \bibnamefont {Afshar}}, \bibinfo
		{author} {\bibfnamefont {B.}~\bibnamefont {Michon}}, \bibinfo {author}
		{\bibfnamefont {F.}~\bibnamefont {Lalibert\'e}}, \bibinfo {author}
		{\bibfnamefont {O.}~\bibnamefont {Cyr-Choini\`ere}}, \bibinfo {author}
		{\bibfnamefont {J.-S.}\ \bibnamefont {Zhou}}, \bibinfo {author}
		{\bibfnamefont {S.}~\bibnamefont {Licciardello}}, \bibinfo {author}
		{\bibfnamefont {S.}~\bibnamefont {Wiedmann}}, \bibinfo {author}
		{\bibfnamefont {N.}~\bibnamefont {Doiron-Leyraud}}, \ and\ \bibinfo {author}
		{\bibfnamefont {L.}~\bibnamefont {Taillefer}},\ }\href {\doibase
		10.1103/PhysRevB.95.224517} {\bibfield  {journal} {\bibinfo  {journal} {Phys.
				Rev. B}\ }\textbf {\bibinfo {volume} {95}},\ \bibinfo {pages} {224517}
		(\bibinfo {year} {2017})}\BibitemShut {NoStop}%
	\bibitem [{\citenamefont {Putzke}\ \emph {et~al.}(2021)\citenamefont {Putzke},
		\citenamefont {Benhabib}, \citenamefont {Tabis}, \citenamefont {Ayres},
		\citenamefont {Wang}, \citenamefont {Malone}, \citenamefont {Licciardello},
		\citenamefont {Lu}, \citenamefont {Kondo}, \citenamefont {Takeuchi},
		\citenamefont {Hussey}, \citenamefont {Cooper},\ and\ \citenamefont
		{Carrington}}]{PutzkeNatPhys2021}%
	\BibitemOpen
	\bibfield  {author} {\bibinfo {author} {\bibfnamefont {C.}~\bibnamefont
			{Putzke}}, \bibinfo {author} {\bibfnamefont {S.}~\bibnamefont {Benhabib}},
		\bibinfo {author} {\bibfnamefont {W.}~\bibnamefont {Tabis}}, \bibinfo
		{author} {\bibfnamefont {J.}~\bibnamefont {Ayres}}, \bibinfo {author}
		{\bibfnamefont {Z.}~\bibnamefont {Wang}}, \bibinfo {author} {\bibfnamefont
			{L.}~\bibnamefont {Malone}}, \bibinfo {author} {\bibfnamefont
			{S.}~\bibnamefont {Licciardello}}, \bibinfo {author} {\bibfnamefont
			{J.}~\bibnamefont {Lu}}, \bibinfo {author} {\bibfnamefont {T.}~\bibnamefont
			{Kondo}}, \bibinfo {author} {\bibfnamefont {T.}~\bibnamefont {Takeuchi}},
		\bibinfo {author} {\bibfnamefont {N.~E.}\ \bibnamefont {Hussey}}, \bibinfo
		{author} {\bibfnamefont {J.~R.}\ \bibnamefont {Cooper}}, \ and\ \bibinfo
		{author} {\bibfnamefont {A.}~\bibnamefont {Carrington}},\ }\href {\doibase
		10.1038/s41567-021-01197-0} {\bibfield  {journal} {\bibinfo  {journal} {Nat.
				Phys.}\ }\textbf {\bibinfo {volume} {17}},\ \bibinfo {pages} {826} (\bibinfo
		{year} {2021})}\BibitemShut {NoStop}%
	\bibitem [{\citenamefont {Ino}\ \emph {et~al.}(2000)\citenamefont {Ino},
		\citenamefont {Kim}, \citenamefont {Nakamura}, \citenamefont {Yoshida},
		\citenamefont {Mizokawa}, \citenamefont {Shen}, \citenamefont {Fujimori},
		\citenamefont {Kakeshita}, \citenamefont {Eisaki},\ and\ \citenamefont
		{Uchida}}]{InoPRB2000}%
	\BibitemOpen
	\bibfield  {author} {\bibinfo {author} {\bibfnamefont {A.}~\bibnamefont
			{Ino}}, \bibinfo {author} {\bibfnamefont {C.}~\bibnamefont {Kim}}, \bibinfo
		{author} {\bibfnamefont {M.}~\bibnamefont {Nakamura}}, \bibinfo {author}
		{\bibfnamefont {T.}~\bibnamefont {Yoshida}}, \bibinfo {author} {\bibfnamefont
			{T.}~\bibnamefont {Mizokawa}}, \bibinfo {author} {\bibfnamefont {Z.-X.}\
			\bibnamefont {Shen}}, \bibinfo {author} {\bibfnamefont {A.}~\bibnamefont
			{Fujimori}}, \bibinfo {author} {\bibfnamefont {T.}~\bibnamefont {Kakeshita}},
		\bibinfo {author} {\bibfnamefont {H.}~\bibnamefont {Eisaki}}, \ and\ \bibinfo
		{author} {\bibfnamefont {S.}~\bibnamefont {Uchida}},\ }\href {\doibase
		10.1103/PhysRevB.62.4137} {\bibfield  {journal} {\bibinfo  {journal} {Phys.
				Rev. B}\ }\textbf {\bibinfo {volume} {62}},\ \bibinfo {pages} {4137}
		(\bibinfo {year} {2000})}\BibitemShut {NoStop}%
	\bibitem [{\citenamefont {Mayr}\ \emph {et~al.}(2006)\citenamefont {Mayr},
		\citenamefont {Alvarez}, \citenamefont {Moreo},\ and\ \citenamefont
		{Dagotto}}]{MayrPRB2006}%
	\BibitemOpen
	\bibfield  {author} {\bibinfo {author} {\bibfnamefont {M.}~\bibnamefont
			{Mayr}}, \bibinfo {author} {\bibfnamefont {G.}~\bibnamefont {Alvarez}},
		\bibinfo {author} {\bibfnamefont {A.}~\bibnamefont {Moreo}}, \ and\ \bibinfo
		{author} {\bibfnamefont {E.}~\bibnamefont {Dagotto}},\ }\href {\doibase
		10.1103/PhysRevB.73.014509} {\bibfield  {journal} {\bibinfo  {journal} {Phys.
				Rev. B}\ }\textbf {\bibinfo {volume} {73}},\ \bibinfo {pages} {014509}
		(\bibinfo {year} {2006})}\BibitemShut {NoStop}%
	\bibitem [{\citenamefont {Capone}\ and\ \citenamefont
		{Kotliar}(2006)}]{CaponePRB2006}%
	\BibitemOpen
	\bibfield  {author} {\bibinfo {author} {\bibfnamefont {M.}~\bibnamefont
			{Capone}}\ and\ \bibinfo {author} {\bibfnamefont {G.}~\bibnamefont
			{Kotliar}},\ }\href {\doibase 10.1103/PhysRevB.74.054513} {\bibfield
		{journal} {\bibinfo  {journal} {Phys. Rev. B}\ }\textbf {\bibinfo {volume}
			{74}},\ \bibinfo {pages} {054513} (\bibinfo {year} {2006})}\BibitemShut
	{NoStop}%
	\bibitem [{\citenamefont {Palczewski}\ \emph {et~al.}(2008)\citenamefont
		{Palczewski}, \citenamefont {Kondo}, \citenamefont {Khasanov}, \citenamefont
		{Kolesnikov}, \citenamefont {Timonina}, \citenamefont {Rotenberg},
		\citenamefont {Ohta}, \citenamefont {Bendounan}, \citenamefont {Sassa},
		\citenamefont {Fedorov}, \citenamefont {Pailh\'es}, \citenamefont
		{Santander-Syro}, \citenamefont {Chang}, \citenamefont {Shi}, \citenamefont
		{Mesot}, \citenamefont {Fretwell},\ and\ \citenamefont
		{Kaminski}}]{PalczewskiPRB2008}%
	\BibitemOpen
	\bibfield  {author} {\bibinfo {author} {\bibfnamefont {A.~D.}\ \bibnamefont
			{Palczewski}}, \bibinfo {author} {\bibfnamefont {T.}~\bibnamefont {Kondo}},
		\bibinfo {author} {\bibfnamefont {R.}~\bibnamefont {Khasanov}}, \bibinfo
		{author} {\bibfnamefont {N.~N.}\ \bibnamefont {Kolesnikov}}, \bibinfo
		{author} {\bibfnamefont {A.~V.}\ \bibnamefont {Timonina}}, \bibinfo {author}
		{\bibfnamefont {E.}~\bibnamefont {Rotenberg}}, \bibinfo {author}
		{\bibfnamefont {T.}~\bibnamefont {Ohta}}, \bibinfo {author} {\bibfnamefont
			{A.}~\bibnamefont {Bendounan}}, \bibinfo {author} {\bibfnamefont
			{Y.}~\bibnamefont {Sassa}}, \bibinfo {author} {\bibfnamefont
			{A.}~\bibnamefont {Fedorov}}, \bibinfo {author} {\bibfnamefont
			{S.}~\bibnamefont {Pailh\'es}}, \bibinfo {author} {\bibfnamefont {A.~F.}\
			\bibnamefont {Santander-Syro}}, \bibinfo {author} {\bibfnamefont
			{J.}~\bibnamefont {Chang}}, \bibinfo {author} {\bibfnamefont
			{M.}~\bibnamefont {Shi}}, \bibinfo {author} {\bibfnamefont {J.}~\bibnamefont
			{Mesot}}, \bibinfo {author} {\bibfnamefont {H.~M.}\ \bibnamefont {Fretwell}},
		\ and\ \bibinfo {author} {\bibfnamefont {A.}~\bibnamefont {Kaminski}},\
	}\href {\doibase 10.1103/PhysRevB.78.054523} {\bibfield  {journal} {\bibinfo
			{journal} {Phys. Rev. B}\ }\textbf {\bibinfo {volume} {78}},\ \bibinfo
		{pages} {054523} (\bibinfo {year} {2008})}\BibitemShut {NoStop}%
	\bibitem [{\citenamefont {Plat\'e}\ \emph {et~al.}(2005)\citenamefont
		{Plat\'e}, \citenamefont {Mottershead}, \citenamefont {Elfimov},
		\citenamefont {Peets}, \citenamefont {Liang}, \citenamefont {Bonn},
		\citenamefont {Hardy}, \citenamefont {Chiuzbaian}, \citenamefont {Falub},
		\citenamefont {Shi}, \citenamefont {Patthey},\ and\ \citenamefont
		{Damascelli}}]{PlatePRL2005}%
	\BibitemOpen
	\bibfield  {author} {\bibinfo {author} {\bibfnamefont {M.}~\bibnamefont
			{Plat\'e}}, \bibinfo {author} {\bibfnamefont {J.~D.~F.}\ \bibnamefont
			{Mottershead}}, \bibinfo {author} {\bibfnamefont {I.~S.}\ \bibnamefont
			{Elfimov}}, \bibinfo {author} {\bibfnamefont {D.~C.}\ \bibnamefont {Peets}},
		\bibinfo {author} {\bibfnamefont {R.}~\bibnamefont {Liang}}, \bibinfo
		{author} {\bibfnamefont {D.~A.}\ \bibnamefont {Bonn}}, \bibinfo {author}
		{\bibfnamefont {W.~N.}\ \bibnamefont {Hardy}}, \bibinfo {author}
		{\bibfnamefont {S.}~\bibnamefont {Chiuzbaian}}, \bibinfo {author}
		{\bibfnamefont {M.}~\bibnamefont {Falub}}, \bibinfo {author} {\bibfnamefont
			{M.}~\bibnamefont {Shi}}, \bibinfo {author} {\bibfnamefont {L.}~\bibnamefont
			{Patthey}}, \ and\ \bibinfo {author} {\bibfnamefont {A.}~\bibnamefont
			{Damascelli}},\ }\href {\doibase 10.1103/PhysRevLett.95.077001} {\bibfield
		{journal} {\bibinfo  {journal} {Phys. Rev. Lett.}\ }\textbf {\bibinfo
			{volume} {95}},\ \bibinfo {pages} {077001} (\bibinfo {year}
		{2005})}\BibitemShut {NoStop}%
	\bibitem [{\citenamefont {Horio}\ \emph {et~al.}(2018)\citenamefont {Horio},
		\citenamefont {Hauser}, \citenamefont {Sassa}, \citenamefont {Mingazheva},
		\citenamefont {Sutter}, \citenamefont {Kramer}, \citenamefont {Cook},
		\citenamefont {Nocerino}, \citenamefont {Forslund}, \citenamefont
		{Tjernberg}, \citenamefont {Kobayashi}, \citenamefont {Chikina},
		\citenamefont {Schr\"oter}, \citenamefont {Krieger}, \citenamefont {Schmitt},
		\citenamefont {Strocov}, \citenamefont {Pyon}, \citenamefont {Takayama},
		\citenamefont {Takagi}, \citenamefont {Lipscombe}, \citenamefont {Hayden},
		\citenamefont {Ishikado}, \citenamefont {Eisaki}, \citenamefont {Neupert},
		\citenamefont {M\aa{}nsson}, \citenamefont {Matt},\ and\ \citenamefont
		{Chang}}]{HorioPRL2018}%
	\BibitemOpen
	\bibfield  {author} {\bibinfo {author} {\bibfnamefont {M.}~\bibnamefont
			{Horio}}, \bibinfo {author} {\bibfnamefont {K.}~\bibnamefont {Hauser}},
		\bibinfo {author} {\bibfnamefont {Y.}~\bibnamefont {Sassa}}, \bibinfo
		{author} {\bibfnamefont {Z.}~\bibnamefont {Mingazheva}}, \bibinfo {author}
		{\bibfnamefont {D.}~\bibnamefont {Sutter}}, \bibinfo {author} {\bibfnamefont
			{K.}~\bibnamefont {Kramer}}, \bibinfo {author} {\bibfnamefont
			{A.}~\bibnamefont {Cook}}, \bibinfo {author} {\bibfnamefont {E.}~\bibnamefont
			{Nocerino}}, \bibinfo {author} {\bibfnamefont {O.~K.}\ \bibnamefont
			{Forslund}}, \bibinfo {author} {\bibfnamefont {O.}~\bibnamefont {Tjernberg}},
		\bibinfo {author} {\bibfnamefont {M.}~\bibnamefont {Kobayashi}}, \bibinfo
		{author} {\bibfnamefont {A.}~\bibnamefont {Chikina}}, \bibinfo {author}
		{\bibfnamefont {N.~B.~M.}\ \bibnamefont {Schr\"oter}}, \bibinfo {author}
		{\bibfnamefont {J.~A.}\ \bibnamefont {Krieger}}, \bibinfo {author}
		{\bibfnamefont {T.}~\bibnamefont {Schmitt}}, \bibinfo {author} {\bibfnamefont
			{V.~N.}\ \bibnamefont {Strocov}}, \bibinfo {author} {\bibfnamefont
			{S.}~\bibnamefont {Pyon}}, \bibinfo {author} {\bibfnamefont {T.}~\bibnamefont
			{Takayama}}, \bibinfo {author} {\bibfnamefont {H.}~\bibnamefont {Takagi}},
		\bibinfo {author} {\bibfnamefont {O.~J.}\ \bibnamefont {Lipscombe}}, \bibinfo
		{author} {\bibfnamefont {S.~M.}\ \bibnamefont {Hayden}}, \bibinfo {author}
		{\bibfnamefont {M.}~\bibnamefont {Ishikado}}, \bibinfo {author}
		{\bibfnamefont {H.}~\bibnamefont {Eisaki}}, \bibinfo {author} {\bibfnamefont
			{T.}~\bibnamefont {Neupert}}, \bibinfo {author} {\bibfnamefont
			{M.}~\bibnamefont {M\aa{}nsson}}, \bibinfo {author} {\bibfnamefont {C.~E.}\
			\bibnamefont {Matt}}, \ and\ \bibinfo {author} {\bibfnamefont
			{J.}~\bibnamefont {Chang}},\ }\href {\doibase 10.1103/PhysRevLett.121.077004}
	{\bibfield  {journal} {\bibinfo  {journal} {Phys. Rev. Lett.}\ }\textbf
		{\bibinfo {volume} {121}},\ \bibinfo {pages} {077004} (\bibinfo {year}
		{2018})}\BibitemShut {NoStop}%
	\bibitem [{\citenamefont {Vishik}\ \emph {et~al.}(2014)\citenamefont {Vishik},
		\citenamefont {Bari\ifmmode \check{s}\else \v{s}\fi{}i\ifmmode~\acute{c}\else
			\'{c}\fi{}}, \citenamefont {Chan}, \citenamefont {Li}, \citenamefont {Xia},
		\citenamefont {Yu}, \citenamefont {Zhao}, \citenamefont {Lee}, \citenamefont
		{Meevasana}, \citenamefont {Devereaux}, \citenamefont {Greven},\ and\
		\citenamefont {Shen}}]{VishikPRB2014}%
	\BibitemOpen
	\bibfield  {author} {\bibinfo {author} {\bibfnamefont {I.~M.}\ \bibnamefont
			{Vishik}}, \bibinfo {author} {\bibfnamefont {N.}~\bibnamefont {Bari\ifmmode
				\check{s}\else \v{s}\fi{}i\ifmmode~\acute{c}\else \'{c}\fi{}}}, \bibinfo
		{author} {\bibfnamefont {M.~K.}\ \bibnamefont {Chan}}, \bibinfo {author}
		{\bibfnamefont {Y.}~\bibnamefont {Li}}, \bibinfo {author} {\bibfnamefont
			{D.~D.}\ \bibnamefont {Xia}}, \bibinfo {author} {\bibfnamefont
			{G.}~\bibnamefont {Yu}}, \bibinfo {author} {\bibfnamefont {X.}~\bibnamefont
			{Zhao}}, \bibinfo {author} {\bibfnamefont {W.~S.}\ \bibnamefont {Lee}},
		\bibinfo {author} {\bibfnamefont {W.}~\bibnamefont {Meevasana}}, \bibinfo
		{author} {\bibfnamefont {T.~P.}\ \bibnamefont {Devereaux}}, \bibinfo {author}
		{\bibfnamefont {M.}~\bibnamefont {Greven}}, \ and\ \bibinfo {author}
		{\bibfnamefont {Z.-X.}\ \bibnamefont {Shen}},\ }\href {\doibase
		10.1103/PhysRevB.89.195141} {\bibfield  {journal} {\bibinfo  {journal} {Phys.
				Rev. B}\ }\textbf {\bibinfo {volume} {89}},\ \bibinfo {pages} {195141}
		(\bibinfo {year} {2014})}\BibitemShut {NoStop}%
	\bibitem [{\citenamefont {Bangura}\ \emph {et~al.}(2010)\citenamefont
		{Bangura}, \citenamefont {Rourke}, \citenamefont {Benseman}, \citenamefont
		{Matusiak}, \citenamefont {Cooper}, \citenamefont {Hussey},\ and\
		\citenamefont {Carrington}}]{BanguraPRB2010}%
	\BibitemOpen
	\bibfield  {author} {\bibinfo {author} {\bibfnamefont {A.~F.}\ \bibnamefont
			{Bangura}}, \bibinfo {author} {\bibfnamefont {P.~M.~C.}\ \bibnamefont
			{Rourke}}, \bibinfo {author} {\bibfnamefont {T.~M.}\ \bibnamefont
			{Benseman}}, \bibinfo {author} {\bibfnamefont {M.}~\bibnamefont {Matusiak}},
		\bibinfo {author} {\bibfnamefont {J.~R.}\ \bibnamefont {Cooper}}, \bibinfo
		{author} {\bibfnamefont {N.~E.}\ \bibnamefont {Hussey}}, \ and\ \bibinfo
		{author} {\bibfnamefont {A.}~\bibnamefont {Carrington}},\ }\href {\doibase
		10.1103/PhysRevB.82.140501} {\bibfield  {journal} {\bibinfo  {journal} {Phys.
				Rev. B}\ }\textbf {\bibinfo {volume} {82}},\ \bibinfo {pages} {140501}
		(\bibinfo {year} {2010})}\BibitemShut {NoStop}%
	\bibitem [{\citenamefont {Culo}\ \emph {et~al.}(2021)\citenamefont {Culo},
		\citenamefont {Duffy}, \citenamefont {Ayres}, \citenamefont {Berben},
		\citenamefont {Hsu}, \citenamefont {Hinlopen}, \citenamefont {Bernath},\ and\
		\citenamefont {Hussey}}]{CuloSciPost2021}%
	\BibitemOpen
	\bibfield  {author} {\bibinfo {author} {\bibfnamefont {M.}~\bibnamefont
			{Culo}}, \bibinfo {author} {\bibfnamefont {C.}~\bibnamefont {Duffy}},
		\bibinfo {author} {\bibfnamefont {J.}~\bibnamefont {Ayres}}, \bibinfo
		{author} {\bibfnamefont {M.}~\bibnamefont {Berben}}, \bibinfo {author}
		{\bibfnamefont {Y.-T.}\ \bibnamefont {Hsu}}, \bibinfo {author} {\bibfnamefont
			{R.~D.~H.}\ \bibnamefont {Hinlopen}}, \bibinfo {author} {\bibfnamefont
			{B.}~\bibnamefont {Bernath}}, \ and\ \bibinfo {author} {\bibfnamefont
			{N.~E.}\ \bibnamefont {Hussey}},\ }\href {\doibase
		10.21468/SciPostPhys.11.1.012} {\bibfield  {journal} {\bibinfo  {journal}
			{SciPost Phys.}\ }\textbf {\bibinfo {volume} {11}},\ \bibinfo {pages} {012}
		(\bibinfo {year} {2021})}\BibitemShut {NoStop}%
	\bibitem [{\citenamefont {Yamamoto}\ \emph {et~al.}(2000)\citenamefont
		{Yamamoto}, \citenamefont {Hu},\ and\ \citenamefont
		{Tajima}}]{YamamotoPRB2000}%
	\BibitemOpen
	\bibfield  {author} {\bibinfo {author} {\bibfnamefont {A.}~\bibnamefont
			{Yamamoto}}, \bibinfo {author} {\bibfnamefont {W.-Z.}\ \bibnamefont {Hu}}, \
		and\ \bibinfo {author} {\bibfnamefont {S.}~\bibnamefont {Tajima}},\ }\href
	{\doibase 10.1103/PhysRevB.63.024504} {\bibfield  {journal} {\bibinfo
			{journal} {Phys. Rev. B}\ }\textbf {\bibinfo {volume} {63}},\ \bibinfo
		{pages} {024504} (\bibinfo {year} {2000})}\BibitemShut {NoStop}%
	\bibitem [{\citenamefont {{Matt}}\ \emph {et~al.}(2018)\citenamefont {{Matt}},
		\citenamefont {{Sutter}}, \citenamefont {{Cook}}, \citenamefont {{Sassa}},
		\citenamefont {{M\aa nsson}}, \citenamefont {{Tjernberg}}, \citenamefont
		{{Das}}, \citenamefont {{Horio}}, \citenamefont {{Destraz}}, \citenamefont
		{{Fatuzzo}}, \citenamefont {{Hauser}}, \citenamefont {{Shi}}, \citenamefont
		{{Kobayashi}}, \citenamefont {{Strocov}}, \citenamefont {{Schmitt}},
		\citenamefont {{Dudin}}, \citenamefont {{Hoesch}}, \citenamefont {{Pyon}},
		\citenamefont {{Takayama}}, \citenamefont {{Takagi}}, \citenamefont
		{{Lipscombe}}, \citenamefont {{Hayden}}, \citenamefont {{Kurosawa}},
		\citenamefont {{Momono}}, \citenamefont {{Oda}}, \citenamefont {{Neupert}},\
		and\ \citenamefont {{Chang}}}]{MattNatCommun2018}%
	\BibitemOpen
	\bibfield  {author} {\bibinfo {author} {\bibfnamefont {C.~E.}\ \bibnamefont
			{{Matt}}}, \bibinfo {author} {\bibfnamefont {D.}~\bibnamefont {{Sutter}}},
		\bibinfo {author} {\bibfnamefont {A.~M.}\ \bibnamefont {{Cook}}}, \bibinfo
		{author} {\bibfnamefont {Y.}~\bibnamefont {{Sassa}}}, \bibinfo {author}
		{\bibfnamefont {M.}~\bibnamefont {{M\aa nsson}}}, \bibinfo {author}
		{\bibfnamefont {O.}~\bibnamefont {{Tjernberg}}}, \bibinfo {author}
		{\bibfnamefont {L.}~\bibnamefont {{Das}}}, \bibinfo {author} {\bibfnamefont
			{M.}~\bibnamefont {{Horio}}}, \bibinfo {author} {\bibfnamefont
			{D.}~\bibnamefont {{Destraz}}}, \bibinfo {author} {\bibfnamefont {C.~G.}\
			\bibnamefont {{Fatuzzo}}}, \bibinfo {author} {\bibfnamefont {K.}~\bibnamefont
			{{Hauser}}}, \bibinfo {author} {\bibfnamefont {M.}~\bibnamefont {{Shi}}},
		\bibinfo {author} {\bibfnamefont {M.}~\bibnamefont {{Kobayashi}}}, \bibinfo
		{author} {\bibfnamefont {V.~N.}\ \bibnamefont {{Strocov}}}, \bibinfo {author}
		{\bibfnamefont {T.}~\bibnamefont {{Schmitt}}}, \bibinfo {author}
		{\bibfnamefont {P.}~\bibnamefont {{Dudin}}}, \bibinfo {author} {\bibfnamefont
			{M.}~\bibnamefont {{Hoesch}}}, \bibinfo {author} {\bibfnamefont
			{S.}~\bibnamefont {{Pyon}}}, \bibinfo {author} {\bibfnamefont
			{T.}~\bibnamefont {{Takayama}}}, \bibinfo {author} {\bibfnamefont
			{H.}~\bibnamefont {{Takagi}}}, \bibinfo {author} {\bibfnamefont {O.~J.}\
			\bibnamefont {{Lipscombe}}}, \bibinfo {author} {\bibfnamefont {S.~M.}\
			\bibnamefont {{Hayden}}}, \bibinfo {author} {\bibfnamefont {T.}~\bibnamefont
			{{Kurosawa}}}, \bibinfo {author} {\bibfnamefont {N.}~\bibnamefont
			{{Momono}}}, \bibinfo {author} {\bibfnamefont {M.}~\bibnamefont {{Oda}}},
		\bibinfo {author} {\bibfnamefont {T.}~\bibnamefont {{Neupert}}}, \ and\
		\bibinfo {author} {\bibfnamefont {J.}~\bibnamefont {{Chang}}},\ }\href
	{\doibase 10.1038/s41467-018-03266-0} {\bibfield  {journal} {\bibinfo
			{journal} {Nat. Commun.}\ }\textbf {\bibinfo {volume} {9}},\ \bibinfo {pages}
		{972} (\bibinfo {year} {2018})}\BibitemShut {NoStop}%
	\bibitem [{\citenamefont {Sakakibara}\ \emph {et~al.}(2010)\citenamefont
		{Sakakibara}, \citenamefont {Usui}, \citenamefont {Kuroki}, \citenamefont
		{Arita},\ and\ \citenamefont {Aoki}}]{SakakibaraPRL2010}%
	\BibitemOpen
	\bibfield  {author} {\bibinfo {author} {\bibfnamefont {H.}~\bibnamefont
			{Sakakibara}}, \bibinfo {author} {\bibfnamefont {H.}~\bibnamefont {Usui}},
		\bibinfo {author} {\bibfnamefont {K.}~\bibnamefont {Kuroki}}, \bibinfo
		{author} {\bibfnamefont {R.}~\bibnamefont {Arita}}, \ and\ \bibinfo {author}
		{\bibfnamefont {H.}~\bibnamefont {Aoki}},\ }\href {\doibase
		10.1103/PhysRevLett.105.057003} {\bibfield  {journal} {\bibinfo  {journal}
			{Phys. Rev. Lett.}\ }\textbf {\bibinfo {volume} {105}},\ \bibinfo {pages}
		{057003} (\bibinfo {year} {2010})}\BibitemShut {NoStop}%
	\bibitem [{\citenamefont {Sakakibara}\ \emph {et~al.}(2012)\citenamefont
		{Sakakibara}, \citenamefont {Usui}, \citenamefont {Kuroki}, \citenamefont
		{Arita},\ and\ \citenamefont {Aoki}}]{SakakibaraPRB2012}%
	\BibitemOpen
	\bibfield  {author} {\bibinfo {author} {\bibfnamefont {H.}~\bibnamefont
			{Sakakibara}}, \bibinfo {author} {\bibfnamefont {H.}~\bibnamefont {Usui}},
		\bibinfo {author} {\bibfnamefont {K.}~\bibnamefont {Kuroki}}, \bibinfo
		{author} {\bibfnamefont {R.}~\bibnamefont {Arita}}, \ and\ \bibinfo {author}
		{\bibfnamefont {H.}~\bibnamefont {Aoki}},\ }\href {\doibase
		10.1103/PhysRevB.85.064501} {\bibfield  {journal} {\bibinfo  {journal} {Phys.
				Rev. B}\ }\textbf {\bibinfo {volume} {85}},\ \bibinfo {pages} {064501}
		(\bibinfo {year} {2012})}\BibitemShut {NoStop}%
	\bibitem [{\citenamefont {Pavarini}\ \emph {et~al.}(2001)\citenamefont
		{Pavarini}, \citenamefont {Dasgupta}, \citenamefont {Saha-Dasgupta},
		\citenamefont {Jepsen},\ and\ \citenamefont {Andersen}}]{PavariniPRL2001}%
	\BibitemOpen
	\bibfield  {author} {\bibinfo {author} {\bibfnamefont {E.}~\bibnamefont
			{Pavarini}}, \bibinfo {author} {\bibfnamefont {I.}~\bibnamefont {Dasgupta}},
		\bibinfo {author} {\bibfnamefont {T.}~\bibnamefont {Saha-Dasgupta}}, \bibinfo
		{author} {\bibfnamefont {O.}~\bibnamefont {Jepsen}}, \ and\ \bibinfo {author}
		{\bibfnamefont {O.~K.}\ \bibnamefont {Andersen}},\ }\href
	{http://link.aps.org/doi/10.1103/PhysRevLett.87.047003} {\bibfield  {journal}
		{\bibinfo  {journal} {Phys. Rev. Lett.}\ }\textbf {\bibinfo {volume} {87}},\
		\bibinfo {pages} {047003} (\bibinfo {year} {2001})}\BibitemShut {NoStop}%
	\bibitem [{\citenamefont {Ohta}\ \emph {et~al.}(1991)\citenamefont {Ohta},
		\citenamefont {Tohyama},\ and\ \citenamefont {Maekawa}}]{OhtaPRB1991}%
	\BibitemOpen
	\bibfield  {author} {\bibinfo {author} {\bibfnamefont {Y.}~\bibnamefont
			{Ohta}}, \bibinfo {author} {\bibfnamefont {T.}~\bibnamefont {Tohyama}}, \
		and\ \bibinfo {author} {\bibfnamefont {S.}~\bibnamefont {Maekawa}},\ }\href
	{\doibase 10.1103/PhysRevB.43.2968} {\bibfield  {journal} {\bibinfo
			{journal} {Phys. Rev. B}\ }\textbf {\bibinfo {volume} {43}},\ \bibinfo
		{pages} {2968} (\bibinfo {year} {1991})}\BibitemShut {NoStop}%
	\bibitem [{\citenamefont {Civelli}\ \emph {et~al.}(2005)\citenamefont
		{Civelli}, \citenamefont {Capone}, \citenamefont {Kancharla}, \citenamefont
		{Parcollet},\ and\ \citenamefont {Kotliar}}]{CivelliPRL2005}%
	\BibitemOpen
	\bibfield  {author} {\bibinfo {author} {\bibfnamefont {M.}~\bibnamefont
			{Civelli}}, \bibinfo {author} {\bibfnamefont {M.}~\bibnamefont {Capone}},
		\bibinfo {author} {\bibfnamefont {S.~S.}\ \bibnamefont {Kancharla}}, \bibinfo
		{author} {\bibfnamefont {O.}~\bibnamefont {Parcollet}}, \ and\ \bibinfo
		{author} {\bibfnamefont {G.}~\bibnamefont {Kotliar}},\ }\href {\doibase
		10.1103/PhysRevLett.95.106402} {\bibfield  {journal} {\bibinfo  {journal}
			{Phys. Rev. Lett.}\ }\textbf {\bibinfo {volume} {95}},\ \bibinfo {pages}
		{106402} (\bibinfo {year} {2005})}\BibitemShut {NoStop}%
	\bibitem [{\citenamefont {Presland}\ \emph {et~al.}(1991)\citenamefont
		{Presland}, \citenamefont {Tallon}, \citenamefont {Buckley}, \citenamefont
		{Liu},\ and\ \citenamefont {Flower}}]{PreslandPhysC1991}%
	\BibitemOpen
	\bibfield  {author} {\bibinfo {author} {\bibfnamefont {M.}~\bibnamefont
			{Presland}}, \bibinfo {author} {\bibfnamefont {J.}~\bibnamefont {Tallon}},
		\bibinfo {author} {\bibfnamefont {R.}~\bibnamefont {Buckley}}, \bibinfo
		{author} {\bibfnamefont {R.}~\bibnamefont {Liu}}, \ and\ \bibinfo {author}
		{\bibfnamefont {N.}~\bibnamefont {Flower}},\ }\href {\doibase
		https://doi.org/10.1016/0921-4534(91)90700-9} {\bibfield  {journal} {\bibinfo
			{journal} {Physica C}\ }\textbf {\bibinfo {volume} {176}},\ \bibinfo {pages}
		{95} (\bibinfo {year} {1991})}\BibitemShut {NoStop}%
	\bibitem [{\citenamefont {Cooper}\ \emph {et~al.}(2009)\citenamefont {Cooper},
		\citenamefont {Wang}, \citenamefont {Vignolle}, \citenamefont {Lipscombe},
		\citenamefont {Hayden}, \citenamefont {Tanabe}, \citenamefont {Adachi},
		\citenamefont {Koike}, \citenamefont {Nohara}, \citenamefont {Takagi},
		\citenamefont {Proust},\ and\ \citenamefont {Hussey}}]{CooperScience2009}%
	\BibitemOpen
	\bibfield  {author} {\bibinfo {author} {\bibfnamefont {R.~A.}\ \bibnamefont
			{Cooper}}, \bibinfo {author} {\bibfnamefont {Y.}~\bibnamefont {Wang}},
		\bibinfo {author} {\bibfnamefont {B.}~\bibnamefont {Vignolle}}, \bibinfo
		{author} {\bibfnamefont {O.~J.}\ \bibnamefont {Lipscombe}}, \bibinfo {author}
		{\bibfnamefont {S.~M.}\ \bibnamefont {Hayden}}, \bibinfo {author}
		{\bibfnamefont {Y.}~\bibnamefont {Tanabe}}, \bibinfo {author} {\bibfnamefont
			{T.}~\bibnamefont {Adachi}}, \bibinfo {author} {\bibfnamefont
			{Y.}~\bibnamefont {Koike}}, \bibinfo {author} {\bibfnamefont
			{M.}~\bibnamefont {Nohara}}, \bibinfo {author} {\bibfnamefont
			{H.}~\bibnamefont {Takagi}}, \bibinfo {author} {\bibfnamefont
			{C.}~\bibnamefont {Proust}}, \ and\ \bibinfo {author} {\bibfnamefont {N.~E.}\
			\bibnamefont {Hussey}},\ }\href {\doibase 10.1126/science.1165015} {\bibfield
		{journal} {\bibinfo  {journal} {Science}\ }\textbf {\bibinfo {volume}
			{323}},\ \bibinfo {pages} {603} (\bibinfo {year} {2009})}\BibitemShut
	{NoStop}%
	\bibitem [{\citenamefont {Yokoyama}\ \emph {et~al.}(2013)\citenamefont
		{Yokoyama}, \citenamefont {Ogata}, \citenamefont {Tanaka}, \citenamefont
		{Kobayashi},\ and\ \citenamefont {Tsuchiura}}]{YokoyamaJPSJ2013}%
	\BibitemOpen
	\bibfield  {author} {\bibinfo {author} {\bibfnamefont {H.}~\bibnamefont
			{Yokoyama}}, \bibinfo {author} {\bibfnamefont {M.}~\bibnamefont {Ogata}},
		\bibinfo {author} {\bibfnamefont {Y.}~\bibnamefont {Tanaka}}, \bibinfo
		{author} {\bibfnamefont {K.}~\bibnamefont {Kobayashi}}, \ and\ \bibinfo
		{author} {\bibfnamefont {H.}~\bibnamefont {Tsuchiura}},\ }\href {\doibase
		10.7566/JPSJ.82.014707} {\bibfield  {journal} {\bibinfo  {journal} {J. Phys.
				Soc. Jpn.}\ }\textbf {\bibinfo {volume} {82}},\ \bibinfo {pages} {014707}
		(\bibinfo {year} {2013})}\BibitemShut {NoStop}%
	\bibitem [{\citenamefont {Watanabe}\ \emph {et~al.}(2023)\citenamefont
		{Watanabe}, \citenamefont {Shirakawa}, \citenamefont {Seki}, \citenamefont
		{Sakakibara}, \citenamefont {Kotani}, \citenamefont {Ikeda},\ and\
		\citenamefont {Yunoki}}]{WatanabeJPCM2023}%
	\BibitemOpen
	\bibfield  {author} {\bibinfo {author} {\bibfnamefont {H.}~\bibnamefont
			{Watanabe}}, \bibinfo {author} {\bibfnamefont {T.}~\bibnamefont {Shirakawa}},
		\bibinfo {author} {\bibfnamefont {K.}~\bibnamefont {Seki}}, \bibinfo {author}
		{\bibfnamefont {H.}~\bibnamefont {Sakakibara}}, \bibinfo {author}
		{\bibfnamefont {T.}~\bibnamefont {Kotani}}, \bibinfo {author} {\bibfnamefont
			{H.}~\bibnamefont {Ikeda}}, \ and\ \bibinfo {author} {\bibfnamefont
			{S.}~\bibnamefont {Yunoki}},\ }\href {\doibase 10.1088/1361-648X/acc0bf}
	{\bibfield  {journal} {\bibinfo  {journal} {J. Phys. Condens. Matter}\
		}\textbf {\bibinfo {volume} {35}},\ \bibinfo {pages} {195601} (\bibinfo
		{year} {2023})}\BibitemShut {NoStop}%
	\bibitem [{\citenamefont {Sawa}\ \emph {et~al.}(1989)\citenamefont {Sawa},
		\citenamefont {Suzuki}, \citenamefont {Watanabe}, \citenamefont {Akimitsu},
		\citenamefont {Matsubara}, \citenamefont {Watabe}, \citenamefont {Uchida},
		\citenamefont {Kokusho}, \citenamefont {Asano}, \citenamefont {Izumi},\ and\
		\citenamefont {Takayama-Muromachi}}]{SawaNature1989}%
	\BibitemOpen
	\bibfield  {author} {\bibinfo {author} {\bibfnamefont {H.}~\bibnamefont
			{Sawa}}, \bibinfo {author} {\bibfnamefont {S.}~\bibnamefont {Suzuki}},
		\bibinfo {author} {\bibfnamefont {M.}~\bibnamefont {Watanabe}}, \bibinfo
		{author} {\bibfnamefont {J.}~\bibnamefont {Akimitsu}}, \bibinfo {author}
		{\bibfnamefont {H.}~\bibnamefont {Matsubara}}, \bibinfo {author}
		{\bibfnamefont {H.}~\bibnamefont {Watabe}}, \bibinfo {author} {\bibfnamefont
			{S.-i.}\ \bibnamefont {Uchida}}, \bibinfo {author} {\bibfnamefont
			{K.}~\bibnamefont {Kokusho}}, \bibinfo {author} {\bibfnamefont
			{H.}~\bibnamefont {Asano}}, \bibinfo {author} {\bibfnamefont
			{F.}~\bibnamefont {Izumi}}, \ and\ \bibinfo {author} {\bibfnamefont
			{E.}~\bibnamefont {Takayama-Muromachi}},\ }\href {\doibase 10.1038/337347a0}
	{\bibfield  {journal} {\bibinfo  {journal} {Nature}\ }\textbf {\bibinfo
			{volume} {337}},\ \bibinfo {pages} {347} (\bibinfo {year}
		{1989})}\BibitemShut {NoStop}%
	\bibitem [{\citenamefont {Radaelli}\ \emph {et~al.}(1994)\citenamefont
		{Radaelli}, \citenamefont {Hinks}, \citenamefont {Mitchell}, \citenamefont
		{Hunter}, \citenamefont {Wagner}, \citenamefont {Dabrowski}, \citenamefont
		{Vandervoort}, \citenamefont {Viswanathan},\ and\ \citenamefont
		{Jorgensen}}]{RadaelliPRB1994_2}%
	\BibitemOpen
	\bibfield  {author} {\bibinfo {author} {\bibfnamefont {P.~G.}\ \bibnamefont
			{Radaelli}}, \bibinfo {author} {\bibfnamefont {D.~G.}\ \bibnamefont {Hinks}},
		\bibinfo {author} {\bibfnamefont {A.~W.}\ \bibnamefont {Mitchell}}, \bibinfo
		{author} {\bibfnamefont {B.~A.}\ \bibnamefont {Hunter}}, \bibinfo {author}
		{\bibfnamefont {J.~L.}\ \bibnamefont {Wagner}}, \bibinfo {author}
		{\bibfnamefont {B.}~\bibnamefont {Dabrowski}}, \bibinfo {author}
		{\bibfnamefont {K.~G.}\ \bibnamefont {Vandervoort}}, \bibinfo {author}
		{\bibfnamefont {H.~K.}\ \bibnamefont {Viswanathan}}, \ and\ \bibinfo {author}
		{\bibfnamefont {J.~D.}\ \bibnamefont {Jorgensen}},\ }\href {\doibase
		10.1103/PhysRevB.49.4163} {\bibfield  {journal} {\bibinfo  {journal} {Phys.
				Rev. B}\ }\textbf {\bibinfo {volume} {49}},\ \bibinfo {pages} {4163}
		(\bibinfo {year} {1994})}\BibitemShut {NoStop}%
	\bibitem [{\citenamefont {Torardi}\ \emph {et~al.}(1988)\citenamefont
		{Torardi}, \citenamefont {Subramanian}, \citenamefont {Calabrese},
		\citenamefont {Gopalakrishnan}, \citenamefont {McCarron}, \citenamefont
		{Morrissey}, \citenamefont {Askew}, \citenamefont {Flippen}, \citenamefont
		{Chowdhry},\ and\ \citenamefont {Sleight}}]{TorardiPRB1988}%
	\BibitemOpen
	\bibfield  {author} {\bibinfo {author} {\bibfnamefont {C.~C.}\ \bibnamefont
			{Torardi}}, \bibinfo {author} {\bibfnamefont {M.~A.}\ \bibnamefont
			{Subramanian}}, \bibinfo {author} {\bibfnamefont {J.~C.}\ \bibnamefont
			{Calabrese}}, \bibinfo {author} {\bibfnamefont {J.}~\bibnamefont
			{Gopalakrishnan}}, \bibinfo {author} {\bibfnamefont {E.~M.}\ \bibnamefont
			{McCarron}}, \bibinfo {author} {\bibfnamefont {K.~J.}\ \bibnamefont
			{Morrissey}}, \bibinfo {author} {\bibfnamefont {T.~R.}\ \bibnamefont
			{Askew}}, \bibinfo {author} {\bibfnamefont {R.~B.}\ \bibnamefont {Flippen}},
		\bibinfo {author} {\bibfnamefont {U.}~\bibnamefont {Chowdhry}}, \ and\
		\bibinfo {author} {\bibfnamefont {A.~W.}\ \bibnamefont {Sleight}},\ }\href
	{\doibase 10.1103/PhysRevB.38.225} {\bibfield  {journal} {\bibinfo  {journal}
			{Phys. Rev. B}\ }\textbf {\bibinfo {volume} {38}},\ \bibinfo {pages} {225}
		(\bibinfo {year} {1988})}\BibitemShut {NoStop}%
	\bibitem [{\citenamefont {Eisaki}\ \emph {et~al.}(2004)\citenamefont {Eisaki},
		\citenamefont {Kaneko}, \citenamefont {Feng}, \citenamefont {Damascelli},
		\citenamefont {Mang}, \citenamefont {Shen}, \citenamefont {Shen},\ and\
		\citenamefont {Greven}}]{EisakiPRB2004}%
	\BibitemOpen
	\bibfield  {author} {\bibinfo {author} {\bibfnamefont {H.}~\bibnamefont
			{Eisaki}}, \bibinfo {author} {\bibfnamefont {N.}~\bibnamefont {Kaneko}},
		\bibinfo {author} {\bibfnamefont {D.~L.}\ \bibnamefont {Feng}}, \bibinfo
		{author} {\bibfnamefont {A.}~\bibnamefont {Damascelli}}, \bibinfo {author}
		{\bibfnamefont {P.~K.}\ \bibnamefont {Mang}}, \bibinfo {author}
		{\bibfnamefont {K.~M.}\ \bibnamefont {Shen}}, \bibinfo {author}
		{\bibfnamefont {Z.-X.}\ \bibnamefont {Shen}}, \ and\ \bibinfo {author}
		{\bibfnamefont {M.}~\bibnamefont {Greven}},\ }\href {\doibase
		10.1103/PhysRevB.69.064512} {\bibfield  {journal} {\bibinfo  {journal} {Phys.
				Rev. B}\ }\textbf {\bibinfo {volume} {69}},\ \bibinfo {pages} {064512}
		(\bibinfo {year} {2004})}\BibitemShut {NoStop}%
\end{thebibliography}

%

\end{document}